\documentclass[sigconf]{acmart}
\AtBeginDocument{%
  }

\copyrightyear{2026}
\acmYear{2026}
\setcopyright{cc}
\setcctype{by-nc-nd}
\acmConference[CHI '26]{Proceedings of the 2026 CHI Conference on Human Factors in Computing Systems}{April 13--17, 2026}{Barcelona, Spain}
\acmBooktitle{Proceedings of the 2026 CHI Conference on Human Factors in Computing Systems (CHI '26), April 13--17, 2026, Barcelona, Spain}
\acmDOI{10.1145/3772318.3791037}
\acmISBN{979-8-4007-2278-3/2026/04}

\usepackage{geometry}
\usepackage{booktabs} 
\usepackage{longtable} 
\usepackage{array}  
\usepackage{xcolor}
\usepackage{subfig}
\usepackage{tabularx}
\usepackage{makecell}
\usepackage{multirow}
\usepackage{acmart-taps}
\begin{document}

\title{\textit{GatheringSense:} AI-Generated Imagery and Embodied Experiences for Understanding Literati Gatherings}


\author{You Zhou}
\affiliation{%
  \institution{
  The Hong Kong University of Science and Technology (Guangzhou)}
  \city{Guangzhou}
  \country{China}}
\email{yzhou785@connect.hkust-gz.edu.cn}

\author{Bingyuan Wang}
\affiliation{%
  \institution{
  The Hong Kong University of Science and Technology (Guangzhou)}
  \city{Guangzhou}
  \country{China}}
\email{bwang667@connect.hkust-gz.edu.cn}

\author{Hongcheng Guo}
\affiliation{%
  \institution{
  The Hong Kong University of Science and Technology (Guangzhou)}
  \city{Guangzhou}
  \country{China}}
\email{hguo277@connect.hkust-gz.edu.cn}

\author{Rui Cao}
\affiliation{%
  \institution{
  Xiamen University Malaysia}
  \city{Sepang}
  \country{Malaysia}}
\email{mec2109494@xmu.edu.my}

\author{Zeyu Wang}
\authornote{Corresponding author.}
\affiliation{%
  \institution{
  The Hong Kong University of Science and Technology (Guangzhou)}
  \city{Guangzhou}
  \country{China}}
\affiliation{%
  \institution{
  The Hong Kong University of Science and Technology}
  \city{Hong Kong}
  \country{Hong Kong}}
\email{zeyuwang@ust.hk}

\renewcommand{\shortauthors}{You Zhou et al.}

\begin{abstract}
Chinese literati gatherings (\textit{Wenren Yaji}), as a situated form of Chinese traditional culture, remain underexplored in depth. Although generative AI supports powerful multimodal generation, current cultural applications largely emphasize aesthetic reproduction and struggle to convey the deeper meanings of cultural rituals and social frameworks. Based on embodied cognition, we propose an AI-driven dual-path framework for cultural understanding, which we instantiate through \textit{GatheringSense}, a literati-gathering experience. We conduct a mixed-methods study ($N=48$) to compare how AI-generated multimodal content and embodied participation complement each other in supporting the understanding of literati gatherings and fostering cultural resonance. Our results show that AI-generated content effectively improves the readability of cultural symbols and initial emotional attraction, yet limitations in physical coherence and micro-level credibility may affect users' satisfaction. In contrast, embodied experience significantly deepens participants' understanding of ritual rules and social roles, and increases their psychological closeness and presence. Based on these findings, we offer empirical evidence and five transferable design implications for generative experience in cultural heritage.
\end{abstract}



\maketitle

\section{Introduction}\label{introduction}
\begin{center}
\textit{``Culture is webs of significance that man himself has spun.''\\\hspace{15em}——Clifford Geertz}
\end{center}

In today's digital era, digital technologies have dramatically increased access to cultural content, yet a fundamental problem remains: seeing culture is not the same as understanding it. Cultural understanding still faces a core challenge: while advanced media can transmit surface-level cultural symbols, they often fail to convey the deeper networks of meaning and the internal logic of cultural practices~\cite{geertz2017interpretation}. This limitation is particularly pronounced for practices in which meaning is distributed across bodily actions, object use, spatial configuration, and interpersonal interaction context, where visual presentation alone is insufficient for deep understanding. Chinese literati gatherings (\textit{Wenren Yaji}) exemplify such complex cultural practices. Since the Wei–Jin period and flourishing through the Tang–Song dynasties, literati gatherings have tightly interwoven poetry, calligraphy, painting, music, and tea ceremony with structured interactions such as turn-taking composition, collaborative creation, spatial orchestration, and ritualized social exchange~\cite{clunas1997pictures,elman2000cultural}.

The emergence of generative AI (GenAI) has created new opportunities for cultural visualization, enabling rapid creation of high-quality images, videos, and multimodal narratives that can make cultural symbols more accessible~\cite{ramesh2021zero,wang2025diffusion}. However, existing research primarily focuses on technical implementation or creative assistance, with limited exploration of how AI-generated content (AIGC) can guide users from passive ``viewing'' to active ``participation'' in cultural understanding~\cite{zhou2024generative, wang2025emotionlens}. Moreover, even visually impressive AI content often struggles to capture procedural rules, embodied rhythms, role relationships, or social atmospheres that are central to cultural practices. 

Meanwhile, immersive technologies have shown potential in cultural heritage applications, yet most focus on tangible cultural heritage reconstruction rather than the structured reproduction of social activity patterns and ritual sequences~\cite{bekele2018survey,champion2022playing}. This overlooks the essentially embodied nature of cultural understanding that cognition emerges through dynamic interactions between body, mind, and environment, not merely through observation~\cite{varela2017embodied,wilson2002six}. For practices like literati gatherings, whose core lies in interaction rhythms, object passing, and mutual acknowledgment, this omission is particularly critical.

These two trends reveal a critical gap: GenAI excels at making culture visible, while immersive and embodied technologies excel at making culture experiential, yet the two are rarely integrated systematically. Many digital cultural experiences still keep users at the level of ``viewing content'' rather than ``entering culture through action.'' This static orientation is especially insufficient for literati gatherings, which center on dynamic human–object–field interactions, and it motivates our introduction of an embodied perspective and a dual-path framework~\cite{la2020drone}.

Based on the embodied cognition theory, we propose an AI-driven dual-path cultural understanding framework. Subsequently, we instantiate this framework through \textit{GatheringSense}, an AI-driven dual-path cultural experience centered around literati gatherings, to explore how a cognitive connection of cultural understanding can be formed between visual recognition and embodied participation, and to raise the following research questions:
\begin{itemize}
    \item \textbf{RQ1:} How does AI-generated multimodal content influence participants' initial understanding of the scenes, activities, and roles in literati gatherings?
    \item \textbf{RQ2:} How do embodied experiences enhance immersion and social presence, and support participants' deeper understanding of the cultural meanings of literati gatherings?
    \item \textbf{RQ3:}  How do these two paths work together to foster participants' cultural resonance with literati gatherings?
\end{itemize}

To answer these questions, we conducted a mixed-methods study ($N=48$). We measured symbol readability, cultural resonance, immersion, and social presence, and collected detailed participants' insights through semi-structured interviews and KANO analysis. The results indicate that AI-generated content significantly enhances participants’ understanding of cultural symbols, with notable improvements in their emotional connection to the concept of literati gatherings. Moreover, image content proved more effective than video in conveying activity intentions and key symbols. Participants also showed a preference for traditional visual styles, Ink-wash (\textit{Xieyi}) and Fine-brush (\textit{Gongbi}), over oil painting or cartoon styles, as the former aligned better with cultural symbols and reduced cognitive load. Furthermore, the embodied experiences contributed to stronger social presence and cultural resonance, and the transition from viewing to participation led to a deeper emotional engagement. Exploratory findings from cross-cultural participants and children suggest that the framework has a certain degree of transferability. Based on the analysis, we further distilled several practice-oriented design implications.

In summary, our main contributions include:
\begin{itemize}
    \item {We propose an AI-driven dual-path framework for cultural understanding, clarifying the cognitive connection between visual interpretation and embodied participation.}
    \item {We conduct a study in a real setting ($N=48$) to show how the two paths jointly support cultural understanding through the stages of ``seeing-perceiving-resonating.''}
    \item {By integrating quantitative and qualitative data with KANO analysis, we derive five transferable design implications for cultural experiences that offer concrete and actionable guidance for creating cultural experiences in digital cultural-heritage contexts.}
\end{itemize}
\section{Background and Concepts}\label{background and concepts}
\subsection{Literati Gathering as a Complex Cultural Practice}\label{sec2.1}

Chinese literati gatherings (\textit{Wenren Yaji}) represent one of the most sophisticated forms of traditional Chinese cultural practice, integrating poetry, calligraphy, painting, music, tea ceremony, and scholarly discourse into highly structured social interactions. These gatherings emerged during the Wei-Jin period (220--420~CE) and flourished throughout the Tang-Song era (618--1279~CE), becoming a defining characteristic of Chinese intellectual culture~\cite{owen2020late}.

Unlike casual social meetings, literati gatherings operate through complex symbolic systems and ritualized interaction patterns. Participants engage in turn-taking poetry composition, collaborative artistic creation, and structured aesthetic evaluation, all orchestrated within carefully designed spatial arrangements that include specific seating orders, object placements, and environmental settings~\cite{hay1994boundaries, clunas2004superfluous}. 

Contemporary attempts to revive these gatherings often struggle with their inherent complexity, frequently failing to capture the underlying social dynamics and cultural meanings beyond surface-level aesthetic reproduction. Therefore, the gathering is not merely an image of ``poetry and paintings'', but rather a complete set of social scenarios centered around ``how to spend this moment together.''

\begin{figure*}[t]
  \centering
    \includegraphics[width=0.9\textwidth]{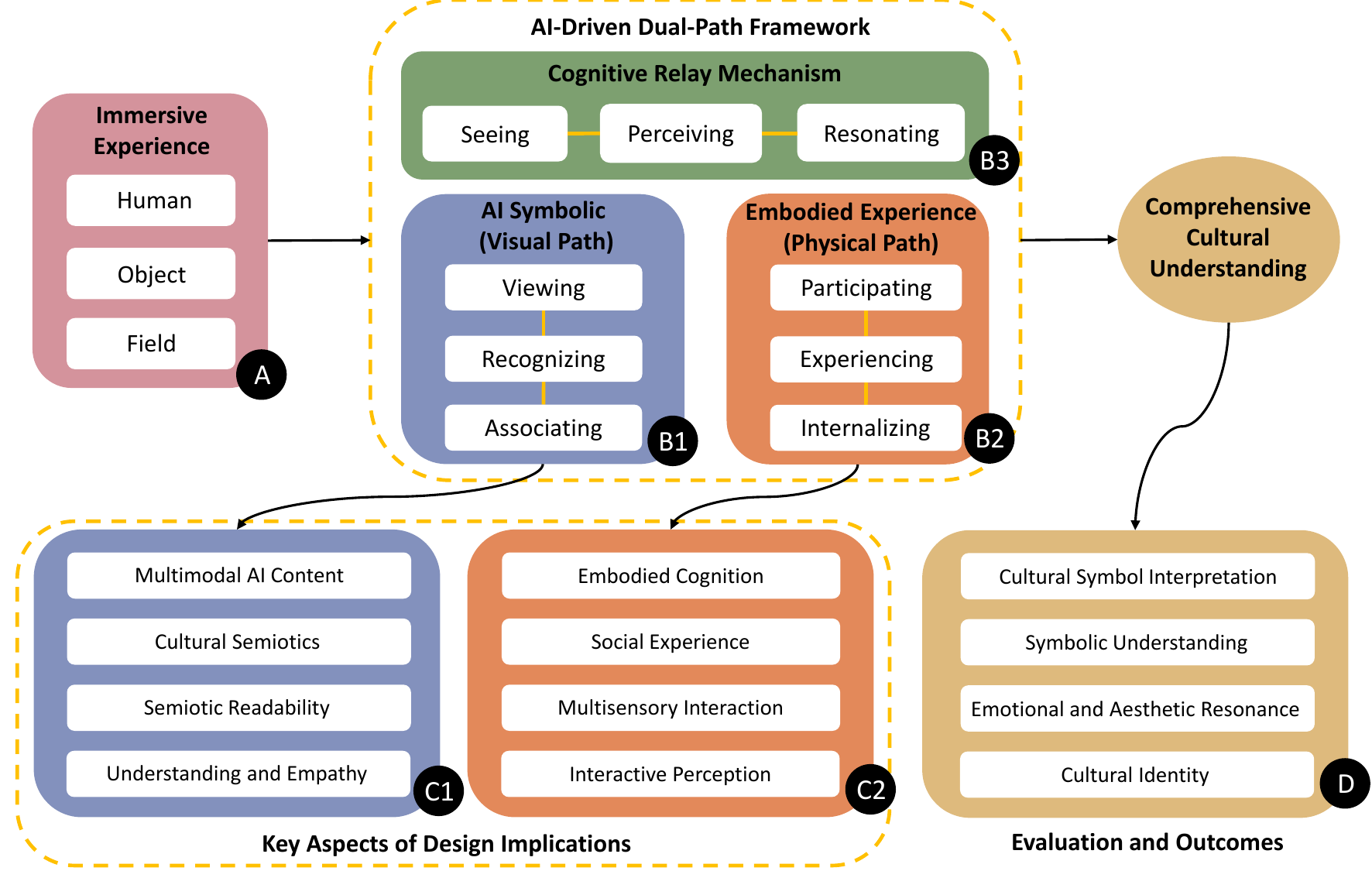}
    \caption{\textbf{AI-driven dual-path framework for cultural understanding.} (A) Immersive setup spans human–object–field. (B1-C1) AI symbolic visual path: AI multimodal content $\rightarrow$ cultural semiotics $\rightarrow$ semiotic readability $\rightarrow$ empathy, priming interpretation and participation. (B2-C2) Embodied experience physical path: embodied cognition + social experience + multisensory/interactive cues, deepening engagement. (B3) Shared cognitive relay mechanism: seeing $\rightarrow$ perceiving $\rightarrow$ resonating (collaboration enhances the handoff). (D) Outcomes: cultural-symbol interpretation, symbolic understanding, emotional/aesthetic resonance, and cultural identity. The two paths are complementary rather than substitutive, jointly driving the understanding and resonance of literati gatherings.}
    \Description{A conceptual diagram illustrating an AI-driven dual-path framework for cultural understanding. The diagram depicts an immersive setup encompassing human, object, and field dimensions. One path represents an AI symbolic visual route, where AI-generated multimodal content supports cultural semiotics, semiotic readability, and empathy, priming interpretation and participation. A second path represents an embodied experience route, combining embodied cognition, social experience, and multisensory interactive cues to deepen engagement. A shared cognitive relay mechanism links seeing, perceiving, and resonating, with collaboration supporting this transition. The framework leads to outcomes including cultural-symbol interpretation, symbolic understanding, emotional and aesthetic resonance, and cultural identity, with the two paths functioning as complementary processes.}
\label{fig:dualpathframework}
\end{figure*}

Drawing from Geertz's concept of ``thick description''~\cite {geertz2017interpretation}, cultural understanding occurs across multiple levels: basic recognition of cultural elements, aesthetic appreciation, understanding of structural rules and social protocols, and grasping underlying cultural values and symbolic significance. Most people encountering unfamiliar cultural practices can readily access the first two levels through observation, but struggle to reach deeper levels without participatory engagement. This limitation is particularly pronounced for ritualized social practices like literati gatherings, where cultural significance lies not just in what is seen, but in how participants interact, create together, and embody cultural values through their actions.

\subsection{From Embodied Cognition Theory to AI-Driven Dual-Path Framework}\label{sec2.2}

Embodied cognition theory provides a crucial theoretical foundation for understanding how people learn and internalize cultural practices~\cite{varela2017embodied, clark1998being}. Rather than treating cognition as purely mental processing, this theory emphasizes that understanding emerges through dynamic interaction between body, mind, and environment. For cultural practices that are inherently embodied, authentic understanding requires more than intellectual comprehension, involving specific gestures, spatial positioning, object manipulation, and coordinated social action. It requires ``embodied cultural knowledge'': the felt sense of appropriate timing, muscle memory of proper gesture, spatial awareness of social positioning, and emotional attunement to group dynamics.

This embodied dimension explains why cultural practices like literati gatherings cannot be fully understood through observation alone. The subtle art of timing in poetry exchange, appropriate gesture when handling tea implements, and spatial choreography of collaborative painting require bodily participation to be truly grasped. This resonates with Chinese aesthetic accounts of literati gatherings as a relational configuration of human–object–field, in which meaning arises from the dynamic alignment of participants, cultural artifacts, and the surrounding environment~\cite{varela2017embodied}. The digitization of cultural heritage has traditionally focused on preservation and presentation, but faces inherent limitations when attempting to transmit embodied cultural practices that exist as dynamic patterns of human interaction rather than static objects.

Building on the above cultural characteristics and theories of embodied cognition, we propose an AI-driven dual-path framework for cultural understanding (Figure~\ref{fig:dualpathframework}), which addresses both symbolic interpretation and embodied participation and enhances the human–object–field perspective:
\begin{itemize}
 \item \textbf{AI symbolic path.} Through AI-generated representations of scenes, activities, and roles (e.g., text, images, and videos), the structure of events becomes easier to recognize and reason about, supporting an initial ``viewing – recognizing – associating'' stage of understanding.
 \item \textbf{Embodied experience path.} Through bodily engagement with cultural activities, such as posture, action sequences, spatial configuration, and coordinated social interaction, participants enact the behavioral logic of the gathering, supporting a deeper ``participating – experiencing – internalizing'' stage of understanding.
\end{itemize}

Together, these two paths form a three-stage ``Seeing – Perceiving – Resonating'' cognitive connection: in the Seeing stage, AI-generated content makes cultural symbols visible and recognizable; in the Perceiving stage, embodied interaction enables experiential learning of underlying rules and practices; in the Resonating stage, participants integrate these experiences into deeper, personally meaningful cultural understanding.

This framework suggests that AI-generated visualization and embodied experience are complementary mechanisms that can work synergistically to support comprehensive cultural understanding, with AI excelling at making cultural symbols visible while embodied interaction enables internalization of cultural patterns and meanings. Having established this theoretical framework, we now examine how existing research has approached the intersection of AI-generated content, embodied interaction, and cultural understanding.
\section{Related Work}\label{related work}

This section summarizes previous work in AI-generated content for cultural heritage, immersive technologies, and embodied interaction for cultural learning, and multimodal cultural understanding and presence.

\subsection{AI-Generated Content for Cultural Heritage}\label{sec3.1}

Recent advances in generative AI have opened new possibilities for cultural heritage visualization and understanding. Early works primarily focused on single-modal content generation, such as text-to-image synthesis for historical reconstruction~\cite{ramesh2021zero, saharia2022photorealistic}. As the field evolved, multimodal generation methods emerged, capable of handling text, images, and videos simultaneously~\cite{singer2022make, ho2022imagen}. In cultural content generation, several studies have explored AI applications in traditional art style simulation and heritage recreation~\cite{lin2024cursive, wang2023naturality}. Zhang et al.~\cite{zhang2025inkspirit} proposed deep learning-based Chinese ink painting generation, while recent work by Zhao et al.~\cite{zhao2025reviving} conducted a comparative study of AI-generated and hand-crafted mural art recreations, revealing insights into the effectiveness of generative AI for cultural heritage restoration. Fu et al.~\cite{fu2024being} explored how GenAI co-creation can enhance engagement and storytelling in cultural heritage dissemination. Recent developments also include tools like Skybox AI for panoramic cultural scene generation~\cite{teixeira2024ai} and methods like SceneDreamer~\cite{chen2023scenedreamer} for direct 3D scene synthesis from textual descriptions. However, existing AI generation methods face significant limitations in cultural applications. They often lack cultural semantic alignment, failing to capture nuanced cultural meanings embedded in traditional practices. Moreover, most approaches focus on static visual content generation without considering the dynamic, social, and ritualistic aspects that characterize complex cultural phenomena like literati gatherings. The generated content frequently suffers from limited cultural authenticity and struggles to convey the aesthetic and symbolic dimensions essential for cultural understanding.

\subsection{Immersive Technologies and Embodied Interaction for Cultural Learning}\label{sec3.2}

Immersive technologies have been increasingly applied to cultural heritage preservation and education. In the realm of Chinese cultural digitization, several notable works have emerged. ShadowPlay2.5D~\cite{zhao2020shadowplay2} leveraged 2.5D sketch strokes for 360-degree video authoring to enhance classical Chinese poetry appreciation. Breaking Barriers for Classical Chinese~\cite{ching2024breaking} explored Tang poetry experiences in virtual reality, demonstrating the potential of Virtual Reality (VR) for cultural education. MagicScroll~\cite{wang2025magicscroll} focused on scroll display representations for different narrative forms. More recently, Li et al.~\cite{li2025ai} designed an AI-driven VR educational game specifically for aesthetic ambience in ancient Chinese poetry, showcasing the integration of AI and immersive technologies for cultural learning. Recent work has also explored intercultural understanding through immersive technologies. Zhao et al.~\cite{zhao2025immersive} developed Immersive Biography, a VR system supporting intercultural empathy and understanding for displaced cultural objects, while Sabie et al.~\cite{sabie2023our} investigated intercultural heritage exchange through augmented reality, demonstrating how AR can facilitate cross-cultural dialogue and understanding. The application of embodied cognition theory in cultural learning has gained attention, suggesting that physical participation and multisensory experiences enable deeper cultural understanding~\cite{wilson2002six, barsalou2008grounded}. In digital cultural heritage, works have explored various embodied interaction modalities, including haptic feedback~\cite{giannopoulos2008effect}, gesture interaction~\cite{roussou2001interplay}, and spatial navigation~\cite{champion2022playing}. Tools like CHER-Ob~\cite{wang2018cher} provide platforms for cultural heritage research by integrating data analysis with interactive visualization. Despite these advances, current immersive cultural systems primarily focus on individual artifact presentation or single-user experiences, lacking consideration for the social dynamics and collective participation that characterize cultural practices like literati gatherings. Moreover, the integration of AI-generated content with embodied interactions remains underexplored, particularly in terms of how physical interactions can enhance cultural presence and aesthetic appreciation in AI-mediated environments.

\begin{figure*}[t]
    \centering
    \includegraphics[width=0.9\linewidth]{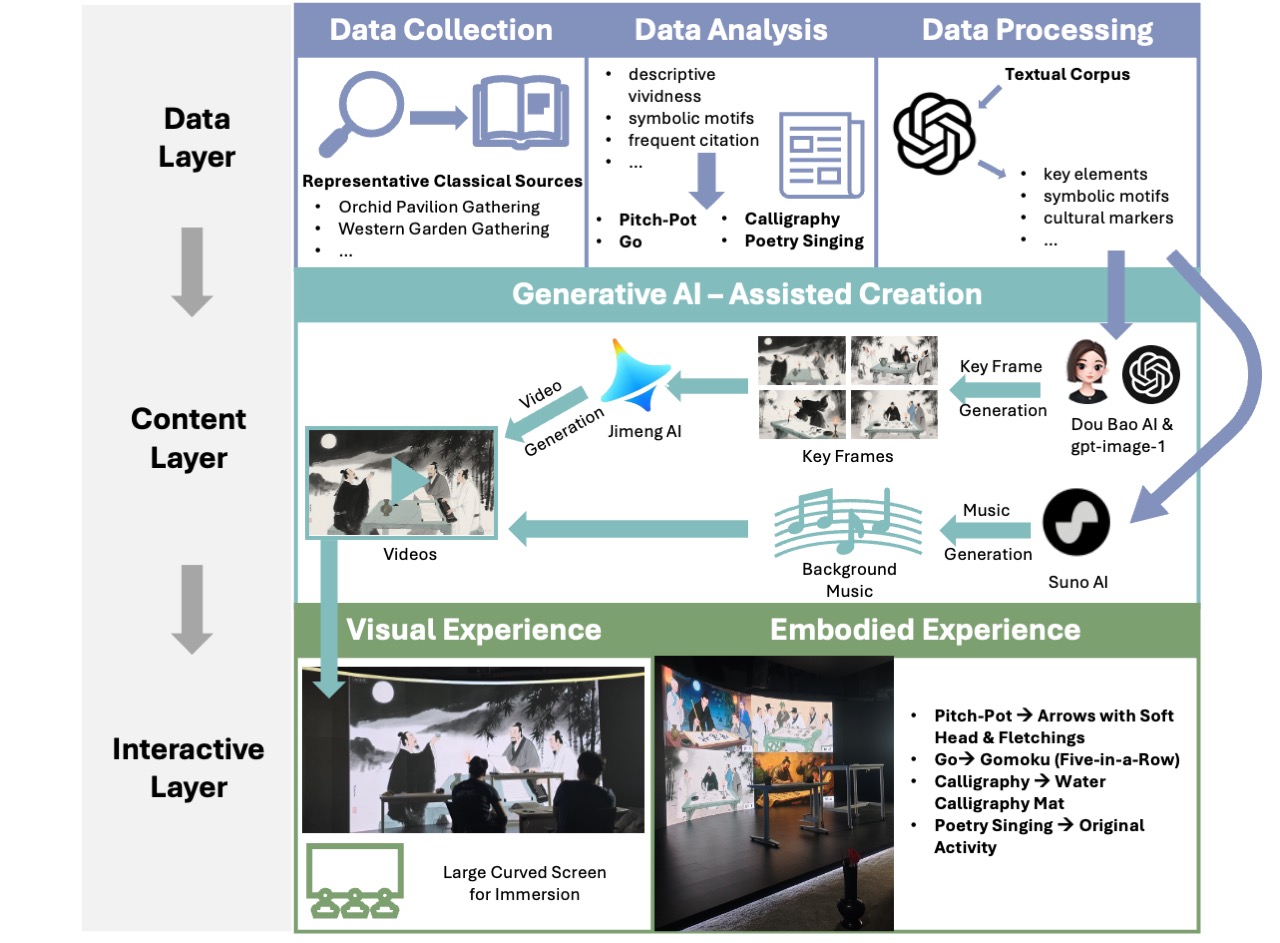}
    \caption{\textbf{Workflow of our methodology.} Our workflow includes three main layers, namely the Data Layer, the Content Layer, and the Interactive Layer. In the Data Layer, we collect, analyze, and process data. In the Content Layer, we utilize multiple generative AIs to create visual content. Finally, we carefully designed visual and embodied experiences, as shown in the Interactive Layer.}
    \Description{A workflow diagram illustrating a study design that integrates AI-based multimodal content with embodied experiences through two complementary paths. The symbolic path consists of AI-generated visual media, including images and short videos displayed on a large curved screen, accompanied by ambient audio and textual descriptions. The embodied path involves physically enacted activities using participants’ bodies, tangible props, spatial arrangement, and live social interaction in a laboratory setting configured as a literati gathering. The workflow is organized into three sequential phases: textual input and semantic extraction, multimodal AI visualization, and embodied activities, with multiple representational styles incorporated throughout the process.}
\label{fig:workflow}
\end{figure*}

\subsection{Multimodal Cultural Understanding and Presence}\label{sec3.3}

Multimodal cultural understanding represents an emerging research direction that combines multiple sensory modalities to enhance cultural comprehension. Existing work mainly concentrates on text-image alignment~\cite{radford2021learning}, cross-modal retrieval~\cite{chen2020uniter}, and multimedia content analysis~\cite{baltruvsaitis2018multimodal}. In cultural applications, some studies have explored semantic matching between poetry and images, associations between traditional music and visual elements, and cross-modal cultural content generation~\cite{zhang2023odorv}. Marti and van der Houwen~\cite{marti2019poetry} demonstrated how poetry can serve as a cross-cultural analysis and sensitizing tool in design, highlighting the potential of literary forms in facilitating cultural understanding. The concept of cultural presence—the subjective feeling of being immersed in a cultural context—extends beyond traditional virtual presence to include dimensions of cultural authenticity, symbolic interpretation, and aesthetic appreciation~\cite{champion2016critical}. Recent work has begun to explore how digital technologies can foster cultural presence. Human-AI collaboration in artistic creation has shown promise, with tools like PromptPaint~\cite{chung2023promptpaint} and VRCopilot~\cite{zhang2024vrcopilot} demonstrating the potential of integrating generative AI with human input through interactive interfaces. LC~\cite{lc2024present} explored participatory generative AI co-created visions as intangible cultural heritage, suggesting new ways of preserving and transmitting cultural knowledge through collaborative AI systems. However, research on multimodal understanding of complex cultural phenomena remains insufficient. Current systems often lack systematic evaluation methods for cultural immersion experiences, particularly for assessing how well digital systems convey cultural meanings and foster aesthetic appreciation. Furthermore, there is limited understanding of how different sensory modalities (visual, auditory, tactile) can be effectively integrated to create coherent cultural experiences, especially when combining AI-generated content with embodied interactions to support the social and ritual dimensions central to cultural practices.

\section{Experience Design}\label{experience design}

Building on the perspective of embodied cognition, we designed a mixed-methods study that combines AI-based media with embodied experiences to examine how the two paths complement each other in supporting understanding of literati gatherings (\textit{Wenren Yaji}). Our measures are organized around three strands of evidence: on the AI side, we focus on symbolic readability and cultural resonance; on the embodied side, we examine immersion, social presence, and changes in symbolic interpretation; and across both, we trace a joint mechanism linking ``viewing–recognizing–associating'' with ``participating–experiencing–internalizing.''

In this study, the term ``AI-based multimodal content'' refers to media generated by AI and integrated in our system. It comprises primarily visual media (i.e., still images and short video clips) presented on a large curved screen, supplemented by controlled ambient audio (music and soundscapes) and contextual textual descriptions. These screen and speaker-based materials constitute the symbolic path. In contrast, the embodied path is implemented through physically enacted activities that engage participants' whole bodies, tangible props, spatial layout, and live social interaction in a lab space configured as a literati gathering environment.

To implement these mechanisms, we structured the study in three phases: (1) textual input and semantic extraction; (2) multimodal AI visualization (text-to-image and text-to-video); and (3) embodied activities in a controlled laboratory environment. Multiple representational styles were incorporated to capture both cultural depth and cognitive diversity. Our entire workflow can be shown in Figure~\ref{fig:workflow}.

\subsection{Textual Corpus and Activity Selection}\label{sec4.1}
\subsubsection{\textbf{Corpus Description.}}\label{sec4.1.1}

To ensure cultural authenticity and textual richness, we constructed a corpus based on Chinese literary history~\cite{chang2010cambridge, Tengwangge}, aesthetics~\cite{lai1973chinese}, and art research~\cite{Xiyuan}, from several classic texts closely related to literati gatherings. We selected five segments that focused on the description of scene imagery, cultural objects, and social scripts. Each text not only provided visual symbols that could be translated by AI, but also provided activity scripts that could be physically re-enacted.
   
\begin{enumerate}
    \item \textbf{\emph{Preface to the Orchid Pavilion Gathering (Lantingji Xu)}} by Wang Xizhi (Eastern Jin). This text records the famous Orchid Pavilion gathering in 353 CE, where scholars composed poems while drinking beside winding streams. The preface is celebrated for its evocation of landscape, conviviality, and transient human emotions. A selected passage was used to represent the archetypal imagery of literati assemblies.  
   
    \item \textbf{\emph{Record of the Elegant Gathering in the Western Garden (Xiyuanyaji Tu Ji)}} by Mi Fu (Northern Song). This account describes Li Gonglin’s painting and the portrayed gathering of eminent literati, including Su Shi and Wang Shen. It emphasizes artistic representation, cultural symbolism, and the refinement of literati presence.  

    \item \textbf{\emph{Collected Records of Jade Mountain (Yushan Mingsheng Ji)}} by Gu Ying (Yuan). 
    This record of a spring gathering in 1348 CE at Jade Mountain Cottage illustrates the integration of painting, music, and poetry among literati friends, with Ni Zan and Yang Weizhen exemplifying the blending of multiple art forms.
   
    \item \textbf{\emph{Preface to the Poems from the Golden Valley (Jingu Shi Xu)}} by Shi Chong (Western Jin). A vivid narrative of feasting and poetic composition at Golden Valley Ravine, highlighting mobility, music, and the performative aspects of gatherings.  
 
    \item \textbf{\emph{Preface to the Pavilion of Prince Teng (Tengwangge Xu)}} by Wang Bo (Tang). Known for its rhetorical flourish, this preface situates literati conviviality within a framework of political authority, hospitality, and the celebration of distinguished guests.  
\end{enumerate}

These sources were translated with reference to existing scholarly conventions and authoritative English renderings where available (e.g., \emph{Lantingji Xu} often translated as \emph{Preface to the Orchid Pavilion Gathering}~\cite{lai1973chinese}). When no standardized version existed, we adopted literal yet stylistically consistent translations, while retaining selected long passages to preserve the cultural imagery. This corpus serves as both the textual comprehension material during the experimental phase and the controlled text input for multimodal AI generation.

\subsubsection{\textbf{Activity Selection.}}\label{sec4.1.2}

\begin{figure}[h]
    \centering
    \includegraphics[width=\linewidth]{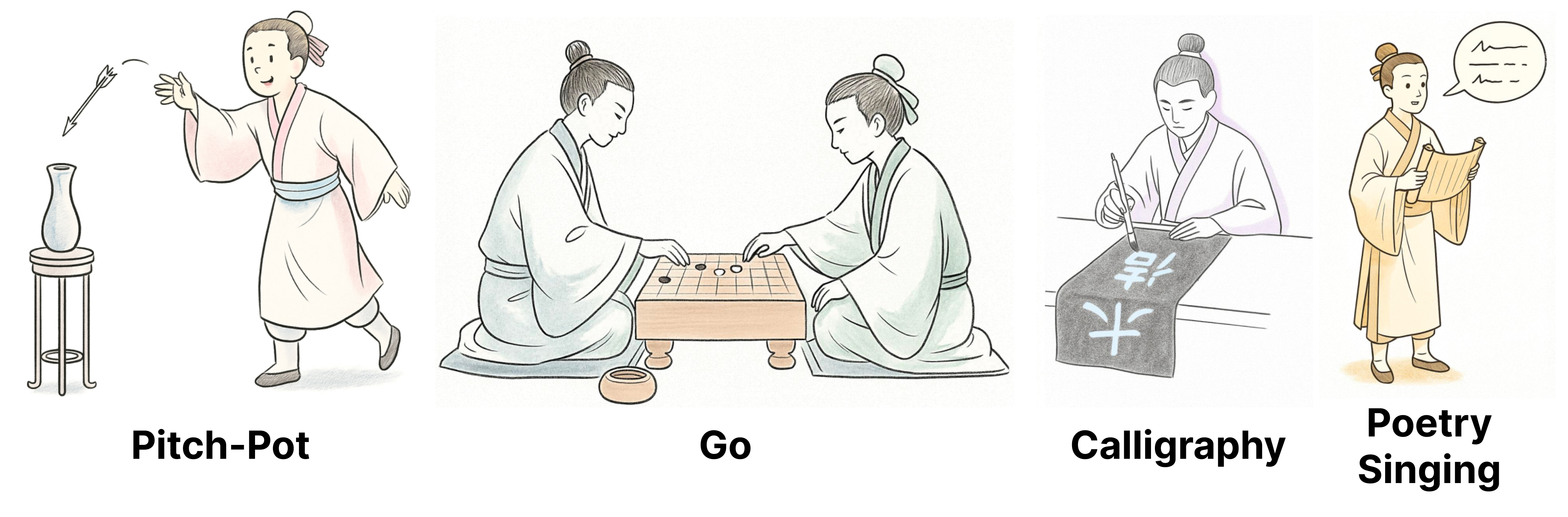}
    \caption{\textbf{Visual demonstration of four core activities.} Pitch-pot, Go, calligraphy, and poetry singing are commonly perceived as the core activities of literati gatherings in historical, literary, and aesthetic contexts.}
    \Description{Images presenting four core activities selected for the study: pitch-pot (touhu), Go, calligraphy, and poetry singing, depicting representative practices in literati gatherings that involve material artifacts, social interaction, and embodied performance.}
    \label{fig:activities}
\end{figure}

Four core activities were selected as experimental foci: pitch-pot (\textit{touhu}), Go, calligraphy, and poetry singing (Figure~\ref{fig:activities}). The selection was guided by three criteria:  
(1) frequent appearance in historical records;  
(2) symbolic value combining material artifacts and social interaction;  
(3) feasibility for embodied enactment in both virtual and physical environments.  
Each activity embodies a different dimension of literati gatherings: ritualized play (pitch-pot), intellectual contest (Go), artistic expression (calligraphy), and collective performance (poetry singing). This diversity helps to detect multiple levels of cultural cognition and symbolic understanding~\cite{Dumoulin2005,lai1973chinese, Liu2008}.

\subsection{AI Visualization Pipeline}\label{sec4.2}
\begin{figure*}[t]
    \centering
    \includegraphics[width=0.7\linewidth]{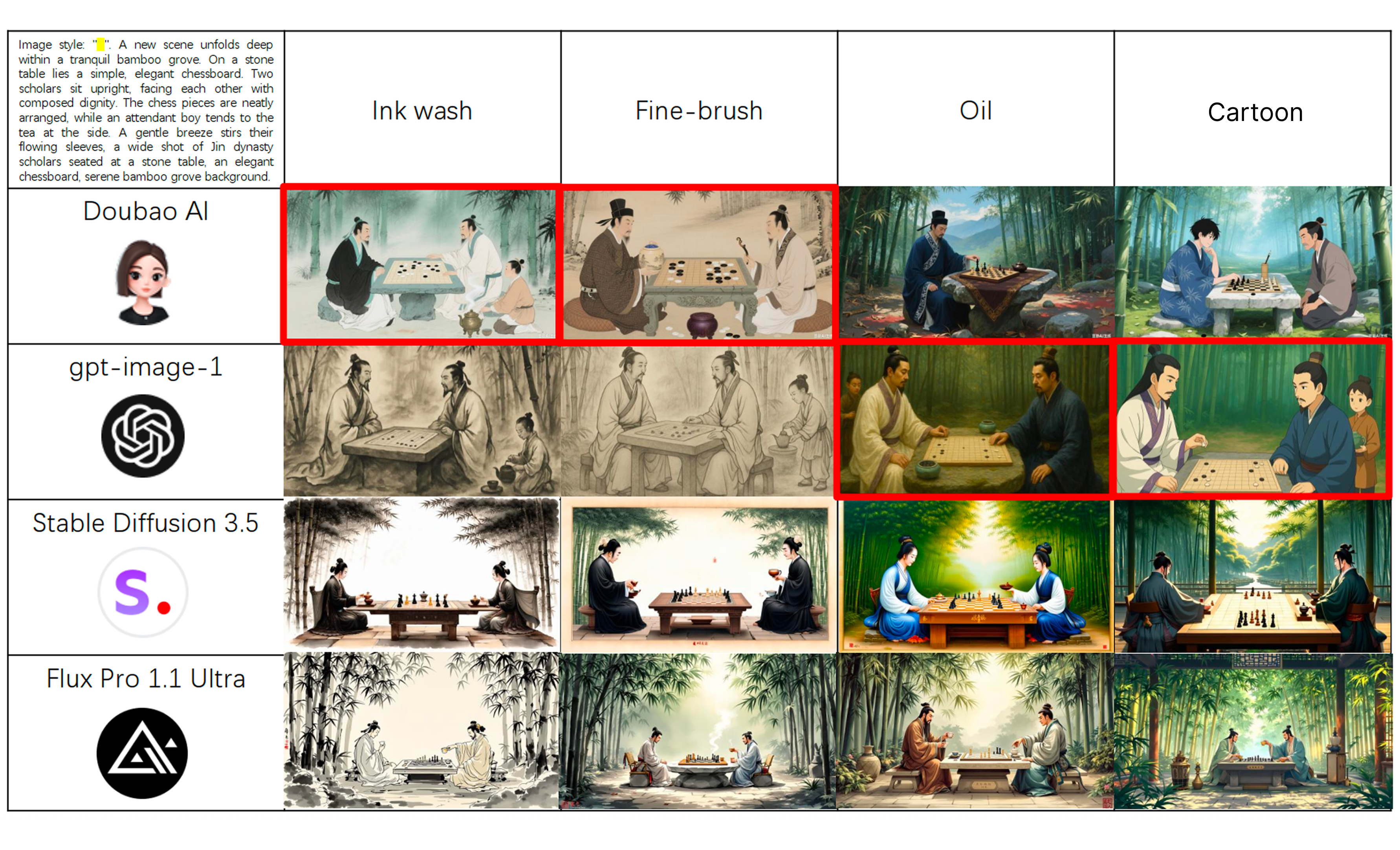}
    \caption{\textbf{Comparison of different models on Go in four styles.} We tested four different AI models by using the same prompt to generate a key frame of Go in four different styles. Dou Bao AI performed best in Ink wash and Fine-brush styles, while gpt-image-1 outperformed in Oil and Cartoon styles. The best results were bounded with red boxes in the figure.}
    \Description{An overall pipeline diagram illustrating the process for generating AI-based multimodal visualizations of four activities. The pipeline proceeds through four sequential stages: text parsing and imagery extraction from textual descriptions; image generation of key frames in multiple visual styles; video composition by interpolating key frames into short clips; and audio and narration integration using music, ambient sounds, and optional narration. The stages are arranged in order to show how textual input is transformed into multimodal visual and audio content while keeping activity content consistent across styles.}
    \label{fig:modelcompare}
\end{figure*}

We designed a structured pipeline to generate AI-based multimodal visualizations of the four activities. The process followed four sequential steps (Figure~\ref{fig:modelcompare}):  

\begin{table}[h]
\centering
\caption{\textbf{AI-generated video materials.} Each video (30--40s) of the four activities was generated in four visual styles, resulting in four clips per activity.}
\label{tab:visualization}
\begin{tabular}{p{2.2cm} c p{2.8cm} c}
\hline
\textbf{Activity} & \textbf{Duration} & \textbf{Visual Styles} & \textbf{Clips} \\
\hline
Pitch-pot & 30--40s & Ink-wash, Fine-brush, Cartoon, Oil & 4 \\
Go & 30--40s & Ink-wash, Fine-brush, Cartoon, Oil & 4 \\
Calligraphy & 30--40s & Ink-wash, Fine-brush, Cartoon, Oil & 4 \\
Poetry Singing & 30--40s & Ink-wash, Fine-brush, Cartoon, Oil & 4 \\
\hline
\textbf{Total} & 2-3 min & & 16 \\
\hline
\end{tabular}
\end{table}

\begin{enumerate}
    \item \textbf{Text Parsing and Imagery Extraction:} Textual descriptions of each activity were processed by GPT-5~\cite{openai2025gpt5},  semantically parsed to extract key elements, symbolic motifs, and cultural markers.  

    \item \textbf{Image Generation (Key Frames):} Using the diffusion or autoregressive-based text-to-image models, we generated 16 key frames in total (four activities $\times$ four styles), capturing critical moments of each activity. The visual styles included: ink-wash (\textit{Xieyi}), fine-brush (\textit{Gongbi}), cartoon, and oil.  Previously, we compared the performance of different models, including gpt-image-1~\cite {openai2025imagegeneration}, Dou Bao AI~\cite{gao2025seedream30technicalreport}, Stable Diffusion 3.5~\cite{stabilityai2024stable3_5}, and Flux Pro 1.1 Ultra~\cite{blackforestlabs2024flux1_1}. As shown in Figure~\ref{fig:modelcompare}, Dou Bao AI had generated the best result on ink wash and fine-brush styles, while gpt-image-1 outperformed on  Oil and Cartoon. The advantages can be partly explained by the data domain and model structures.

    \item \textbf{Video Composition:} The generated key frames were interpolated using frame synthesis algorithms to form smooth video clips of 30--40 seconds each, yielding 16 videos in total (Table~\ref{tab:visualization}). Specifically, we selected Jimeng 2.0 AI~\cite{gao2025seedream30technicalreport} to interpolate key frames. Several insights were collected during the generation. For instance, AI struggled to accurately represent pitch-pot, an activity that is both culturally unfamiliar and physically dynamic. Subsequently, the video of the pitch-pot was poorly appreciated in later studies.

   \item\textbf{Audio and Narration Integration:} Soundscapes were layered onto the videos, including \textit{guqin} and \textit{xiao flute} music, ambient natural sounds, and optional poetic narration. We designed this music with Suno~\cite{suno2025v4_5}, the state-of-the-art AI assistant for music creation. The goal was to reinforce immersion and multimodal coherence.  
\end{enumerate}

This design allowed us to manipulate representational style while keeping activity content constant, thus isolating the effects of visual modality on symbolic readability and cultural resonance.

\subsection{Embodied Experience Setup}\label{sec4.3}
\paragraph{\textbf{Props and Substitutes.}}\label{sec4.3.1}

To translate the virtual activities into embodied enactments, we prepared culturally inspiring, while safe substitutes. For calligraphy, participants used brush pens and imitation paper scrolls. For pitch-pot, lightweight arrows with soft heads and fletchings, and steady ceramic pots were provided. For Go, we settled on Gomoku (Five-in-a-Row) as the duration was considered. For poetry singing, imitation scrolls and slips were prepared. These materials retained symbolic fidelity while ensuring practical safety in the laboratory.

\begin{figure*}[t]
    \centering
    \includegraphics[width=0.8\linewidth]{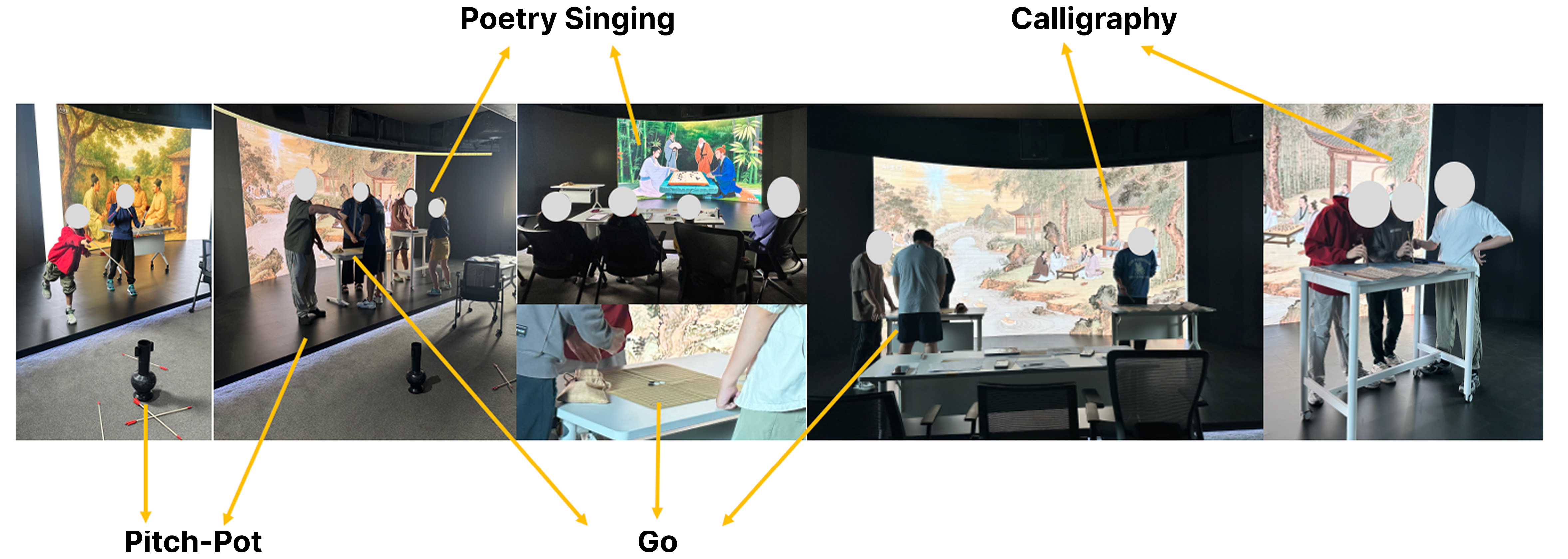}
    \caption{\textbf{Embodied literati gathering in a structured social frame.} Four representative activities were staged within an immersive space, with participants experiencing each one within the social framework of triads.}
    \Description{A diagram illustrating the embodied experimental setup, including culturally inspired props, spatial arrangement, and immersive environment. The figure shows substitute materials used for different activities, such as brush pens and scrolls for calligraphy, lightweight arrows and pots for pitch-pot, a tabletop setup for Go, and scrolls for poetry singing. The laboratory space is arranged with a central table, peripheral activity zones, circular or semi-circular seating, and a large curved screen for media playback. Ambient lighting, background projections, and soundscapes are depicted to convey an immersive environment supporting embodied participation and social interaction.}
    \label{fig:Embodied}
\end{figure*}

\begin{table*}[htbp]
    \centering
    \caption{\textbf{Participant group classification and related information.} This table presents the four experimental groups: Chinese adult forward sequence (G1, $N=23$), Chinese adult reverse sequence (G2, $N=15$), cross-cultural adult (G3, $N=4$), and Chinese children (G4, $N=6$), along with other demographic statistics. Among them, G1 and G2 are main groups for all analyses, G3 and G4 are supplementary groups for only exploratory findings.}
    \label{tab:Participant Demographics}   
    \begin{tabular}{cccccccc}
      \hline
      Group ID & Group Name & Culture & Path Order & $N$ & \makecell[c]{Gender \\(Male/Female)} & \makecell[c]{Education \\(Bachelor \\or higher\%)} & \makecell[c]{AI Media \\Experience \\$M\pm SD$}\\
      \hline
      \multicolumn{8}{c}{\textit{Main Groups}} \\
      \hline
        G1 & CN Forward & Chinese & AI to Embodied & 23 & 13/10 & 95.7\% & $4.65\pm 1.19$ \\
        G2 & CN Reverse & Chinese & Embodied to AI & 15 & 9/6 & 100.0\% & $4.60\pm 0.83$ \\
      \hline
      \multicolumn{8}{c}{\textit{Supplementary Groups}} \\
      \hline
        G3 & Non-CN & English & AI to Embodied & 4 & 3/1 & 75.0\% & $4.00\pm 1.15$ \\
        G4 & Children & Chinese & AI to Embodied & 6 & 1/5 & 0.0\% & $4.00\pm 1.67$ \\
        \bottomrule
    \end{tabular}
\end{table*}

\paragraph{\textbf{Spatial Arrangement.}}\label{sec4.3.2}
The laboratory was transformed into an immersive space simulating a literati gathering. A central table served as the focal site for calligraphy and Go, while peripheral standing zones were designated for pitch-pot. A huge curved screen was used to play media materials, creating an immersive atmosphere. Seating was arranged in circular or semi-circular patterns to foster turn-taking, visual contact, and social acknowledgment. Movement paths were deliberately structured to guide the passing of props (e.g., arrows and scrolls), reinforcing ritualized interaction. Soft lighting and projected backgrounds subtly supported the cultural ambience.

\paragraph{\textbf{Immersive Environment.}}\label{sec4.3.3}
The embodied experience was framed as a participatory performance, alternating between observation and active involvement. Ambient music (\textit{guqin and xiao flute}) and natural soundscapes created an atmospheric backdrop. Participants rotated roles (performer and observer) to reproduce the rhythm of traditional gatherings. This design highlighted immersion, social co-presence, and symbolic interpretation, allowing participants to shift from cognitive recognition to embodied internalization (Figure~\ref{fig:Embodied}).

\begin{figure*}[th]
    \centering
    \includegraphics[width=0.9\linewidth]{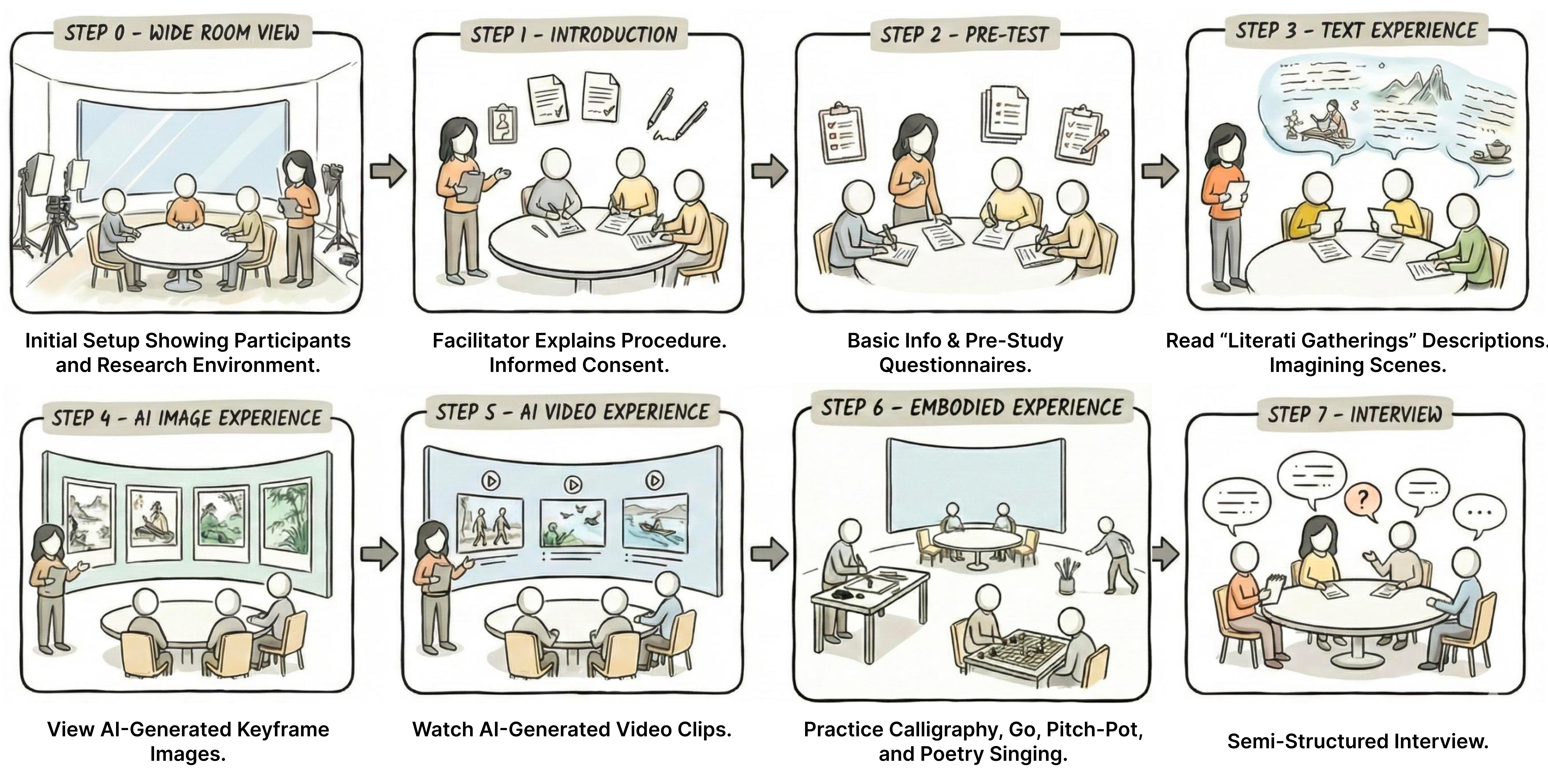}
    \caption{\textbf{Example experience procedure of user study.} All steps are included and arranged according to a complete forward order condition.}
    \Description{A procedural diagram illustrating the user study workflow conducted in a laboratory transformed into a literati gathering space. The figure shows the sequence of experimental steps and group conditions, including forward and reverse orders of AI-media viewing and embodied activities. The procedure includes arrival and social warm-up, consent and pre-test questionnaires, text association, AI image and video viewing with questionnaires, embodied participation in pitch-pot, calligraphy, Go, and poetry singing, followed by post-test measures, and concluding semi-structured interviews and group discussion. Variations across groups are indicated to show differences in step order and the inclusion or omission of selected tasks.}
    \label{fig:Procedure}
\end{figure*}

\section{Method}\label{method}
\subsection{Pilot Test}\label{sec5.1}

Before the user study, we consulted with three experts on literati gatherings to verify the narrative content for accuracy and balance. We then conducted a pilot study with them to assess the end-to-end feasibility and the clarity of our instruments (survey and interview questions). Based on observations and debriefs, we refined the protocol by adding a short tutorial to facilitate the association task and by streamlining the timing of stimuli to improve procedural flow. The duration of the finalized session was kept under 60 minutes, and the participants demonstrated a good comprehension of the rating elements.

\subsection{Participants}\label{sec5.2}

We recruited 48 participants through campus and social media, including 42 adults and 6 children. Among the adults, 25 were male, and 17 were female, with most (approximately 92.86\%) aged between 18 and 30; the children were between 7 and 10. Most adults held a bachelor's or master's degree (about 95.24\% with undergraduate education or above), with a few pursuing or holding a doctoral degree; all children were primary school students. In terms of language and cultural background, the sample included 38 participants with a Chinese cultural background and 4 participants from non-Chinese cultural backgrounds.

Participants were divided into four groups (Table~\ref{tab:Participant Demographics}): Chinese adults in the forward order (G1, $N=23$), Chinese adults in the reverse order (G2, $N=15$), cross-cultural adults (G3, $N=4$), and Chinese children (G4, $N=6$). Note that only G1 and G2 served as the \textbf{main} experimental groups and were included in all subsequent quantitative and qualitative analyses, while G3 and G4 were retained as \textbf{supplementary} groups for exploratory findings. Most participants reported high exposure to AI-generated images and videos (AIGC experience: $M=4.57, SD=1.06$), but low familiarity with literati gatherings (familiarity: $M=3.24, SD=1.21$), showing a typical pattern of being ``technologically familiar but culturally unfamiliar,'' which provided an appropriate basis for examining the dual-path framework under limited cultural prior knowledge. Children were generally familiar with cartoons and gamified media and had some indirect awareness of AI, but had almost no understanding of traditional cultural concepts, such as literati gatherings. Detailed demographic statistics for each group are provided in Appendix~\ref{section:complete participant demographics}.\\

\subsection{Procedure}\label{sec5.3}
The study was conducted in a laboratory that was temporarily transformed into a literati gathering space. Each session lasted approximately 60 minutes, with 3--4 participants completing the full procedure together.

Figure~\ref{fig:Procedure} displays an example experience procedure of our user study. Before the experiment began, participants entered their pre-assigned group condition (G1--G4 in Table~\ref{tab:Participant Demographics}). G1 experienced the dual-path cultural experience in a \textbf{forward} order—first the AI-media phase and then the embodied phase (Steps 0--7). Whereas participants in G2 experienced a \textbf{reverse} order, starting with the embodied phase, followed by the AI-media phase (Steps 0--3, 6, 4--5, 7). This arrangement keeps the content of both paths constant while allowing us to examine whether path order influences key measures. Participants in G3 followed the same forward procedure as G1. G4 followed a simplified forward procedure, in which text association (Step 3), semi-structured interviews (Step 7), and the KANO questionnaires were omitted. As mentioned above, the results of supplementary groups (G3 and G4) are reported separately as exploratory findings.

The detailed procedures for all steps are outlined as follows, using the forward order condition as an example:

\textbf{Steps 0--1: Invitation Framing and Arrival Warm-up.} Before the experiment, we introduced the background of this activity to the participants by saying ``You are invited to attend a small literati gathering.'' We explained the basic concept of the literary and artistic gathering and the general arrangement of this experience. We also invited three participants to introduce themselves one by one and have a brief conversation. This warm-up session was designed to create a social context for a gathering, reduce the sense of strangeness, and enable the participants to enter the atmosphere of the gathering naturally. No formal data collection was conducted in this step.

\textbf{Step 2: Consent and Pre-Test Questionnaire.}
Next, participants were presented with an informed consent form to ensure they understood the study's requirements, including agreeing to video, audio, and image recordings throughout the experiment. After consent, participants completed a pre-test questionnaire, which assessed their familiarity with literati gatherings, understanding of AI-generated images, and expectations for immersive scenarios. Data collection for this phase was managed through the online survey system, with audio recording. This was followed by demographic data collection, cultural contact frequency, and the initial emotional assessments using the Inclusion of Other in the Self (IOS) scale and the Self-Assessment Manikin (SAM) scale, which were administered pre-experiment \cite{Aron1992, Lang1980}.

\textbf{Step 3: Text Association.}
In the third step, participants completed a text association task. They were given short textual descriptions of literati gatherings and asked to write down associated keywords, focusing on symbolic elements (e.g., objects, roles, spatial cues) and other related impressions.

\textbf{Steps 4--5: AI Content Viewing.}
In Step 4, participants viewed a series of 16 key frame images, divided into four types of gathering activities and four visual styles. They were then prompted to complete the Film-IEQ agency scale, SAM\_img, and cultural resonance measures~\cite{Reinecke2014}. In Step 5, participants watched eight video clips, which presented four types of activities across two visual styles. These videos were balanced using a Latin square distribution, ensuring that the music and content remained identical, while only visual styles varied. After viewing the videos, participants completed another Film-IEQ agency, SAM\_vid, and cultural resonance questionnaires. 

\textbf{Step 6: Embodied Experience and Post-Test.}
The sixth step involved an embodied experience where participants actively engaged in four activities typical of a literati gathering: pitch-pot, calligraphy, Go, and poetry singing. These activities were performed within a mixed environment of curved screens and physical props, providing an immersive setting. During the activities, participants observed the correspondence between their actions and AI-generated visual content, reflecting on how the digital representations aligned with traditional practices. Participants were asked whether they understood the spiritual core of the literati gatherings through AI-generated content and were invited to reflect on how these activities might look in contemporary settings. Upon completion of these activities, participants filled out the post-test questionnaire on an online Questionnaire platform, assessing changes in cultural understanding, aesthetic experience, and emotional responses using the IOS, SAM, and the ITC-Sense of Presence Inventory (SOPI) scales~\cite{Lessiter2001}.

\textbf{Step 7: Semi-Structured Interviews and Group Discussion.} 
The final step consisted of semi-structured interviews and group discussions, where participants shared their interpretations of the activities, focusing on the ritualistic significance and the co-creation process they experienced. The discussions also allowed for comparison between the immersive experience and the video viewing, enabling participants to provide feedback on their interpretations and experiences. During this phase, the KANO model was used to capture participants' preferences and suggestions for the immersive experience scenarios, such as the explicitness of visual symbols, the presence of turn-taking prompts, and the use of spatialized soundscapes~\cite{sauerwein1996kano}.

\subsection{Measures}\label{sec5.4}
Following the assessment framework adopted in prior studies~\cite{zhao2025immersive, zhu2025exploring}, we employed both qualitative and quantitative measures to assess key aspects of user experience, primarily focusing on cultural resonance and embodied engagement, while also including presence, affective state, and cognitive change.

\subsubsection{\textbf{Quantitative Measures}}\label{sec5.4.1}

\textbf{Immersive Experience Questionnaire for Film and TV (Film-IEQ) Proxy.}
Adapted from the Film-IEQ~\cite{Reinecke2014}, this scale was used to assess participants' comprehension and involvement when viewing AI-generated images and videos. We focused on items related to behavioral intention understanding and symbolic readability, and administered the scale separately after the image block and the video block to compare how the two media types differ in ``making sense'' and ``feeling involved'' (RQ1).

\textbf{Cultural Resonance Scale (CR).}
We designed a Cultural Resonance Scale that evaluates participants' emotional connection, meaning-making, and aesthetic interest toward literati gatherings across three dimensions: values, rituals, and aesthetics. CR was measured after the image block, after the video block, and after the overall experience, allowing us to trace how resonance develops from AI media to embodied activities (RQ1, RQ2).

\textbf{ITC-Sense of Presence Inventory (ITC-SOPI), Selected Items.}
Based on the ITC-SOPI~\cite{Lessiter2001}, we selected and adapted items to assess spatial presence, engagement, and naturalness during the embodied phase. The scale was administered only after the embodied activities, to examine how the embodied path contributes to a sense of ``being there'' and understanding the social frame of the gathering (RQ2).

\textbf{Inclusion of Other in the Self Scale (IOS).}
We used the standard IOS pictorial scale \cite{Aron1992} to measure participants’ perceived closeness to ``literati gatherings'' as a cultural object. IOS was administered before and after the full experience. We focus on the change score ($\Delta$IOS) as an indicator of shifts in perceived cultural closeness across the two paths (RQ3).

\textbf{Self-Assessment Manikin (SAM).}
The SAM scale captures three affective dimensions—Valence, Arousal, and Dominance~\cite{Lang1980}. We administered SAM at four time points: pre-test, after images, after videos, and after the overall experience. This allows us to map how emotional states evolve as participants move along the two paths from viewing to participation (RQ3).

\textbf{KANO Model Questionnaire.} We constructed paired questions following the KANO model~\cite{sauerwein1996kano} to probe user needs for future immersive experiences, focusing on design features such as multisensory atmosphere, ritual participation, personalized roles, guidance for cultural knowledge, and AI feedback. Each feature was classified into ``must-be,'' ``one-dimensional,'' ``attractive,'' or ``indifferent'' categories, and customer satisfaction ($CS^+$/$CS^-$) coefficients were computed. These results primarily inform design implications derived from RQ3.

\subsubsection{\textbf{Qualitative Indicators}}\label{sec5.4.2}
\textbf{Association keywords.}
After reading the textual corpus, participants completed a ``literati gathering association'' task in which they wrote down words related to characters, actions, objects, scenes, and emotions. We analyzed these association keywords in terms of frequency and semantic categories to understand how participants conceptualized literati gatherings and how this changed across phases.

\textbf{Semi-structured interviews.}
At the end of the experiment, we conducted semi-structured interviews with participants in G1, G2, and G3, focusing on how they experienced and interpreted the AI-symbolic and embodied experience paths, how they understood the rituals and social dynamics of the gatherings, and how they perceived the difference between ``viewing'' and ``participating.'' These qualitative materials were later used for thematic analysis and support the qualitative results reported in Section~\ref{sec6.2}.

\subsection{Data Analysis}\label{sec5.5}
\subsubsection{\textbf{Quantitative Data}}\label{sec5.5.1}

For quantitative analysis, we focused on the core experimental groups (G1+G2) and employed Linear Mixed Models (LMMs), treating participants as random intercepts and modeling phase and style as fixed effects, with cultural contact and prior immersion experience included as covariates~\cite{Baayen2008}. Planned contrasts between key phase pairs (e.g., pre (study) vs. post (study), (after) image vs. (after) video) were conducted to assess changes over time along the dual-path process. For each model, we reported test statistics for main effects and planned contrasts ($F$ or $t$), degrees of freedom (dof), p-values, and corresponding effect sizes (e.g., $\eta^2_p$ or Cohen's $d$). Where appropriate, Tukey corrections were applied for multiple comparisons.

The KANO analysis was conducted to identify user preferences for the design features of the immersive experience. The category proportions and CS coefficients were computed using the KANO model~\cite{sauerwein1996kano}, and the results were visualized with scatter plots. Chi-square and permutation tests were used to compare differences between categories, and False Discovery Rate correction was applied to adjust for multiple comparisons.

Because of the small sample sizes in the cross-cultural group (G3) and children's group (G4), these groups were not included in the LMMs. Instead, we conducted within-group paired-sample $t$-tests or corresponding non-parametric tests to explore pre–post differences. These results were treated as exploratory evidence regarding the dual-path framework in cross-cultural and cross-age contexts.

\subsubsection{\textbf{Qualitative Data}}\label{sec5.5.2}

For qualitative analysis, we transcribed all association keywords and semi-structured interviews and conducted a reflexive thematic analysis~\cite{Braun2006} to identify key themes and experiential patterns across participants. Two researchers first read through the association lists and interview transcripts in MaxQDA and performed open coding on relevant segments. Through iterative discussion, we merged similar codes into higher-level themes and repeatedly returned to the raw data to refine theme boundaries so that they accurately reflected participants' understanding and experience along the dual paths. Cross-cultural cases were marked as a separate subset during coding to support differentiated presentation in the results.

Finally, the mixed data were triangulated, where discrepancies between quantitative and qualitative findings were explored through explanatory hypotheses and boundary conditions to provide a deeper understanding of the participants' experiences.
\section{Results}\label{results}
In this section, we report quantitative and qualitative findings around RQ1--RQ3, grounded in the AI-driven dual-path framework for cultural understanding. Unless otherwise specified, the main statistical analyses in Section~\ref{sec6.1} are based on the Chinese sample G1+G2 ($N=38$). Qualitative trends and findings from cross-cultural individuals (G3) and children (G4) are presented in an exploratory manner in Section~\ref{sec6.2.3}.

\subsection{Quantitative Findings}\label{sec6.1}
\subsubsection{\textbf{Interpretability of AIGC: Media Format and Visual Style}}\label{sec6.1.1}
\begin{figure}[h]
    \centering
    \includegraphics[width=0.37\textwidth]{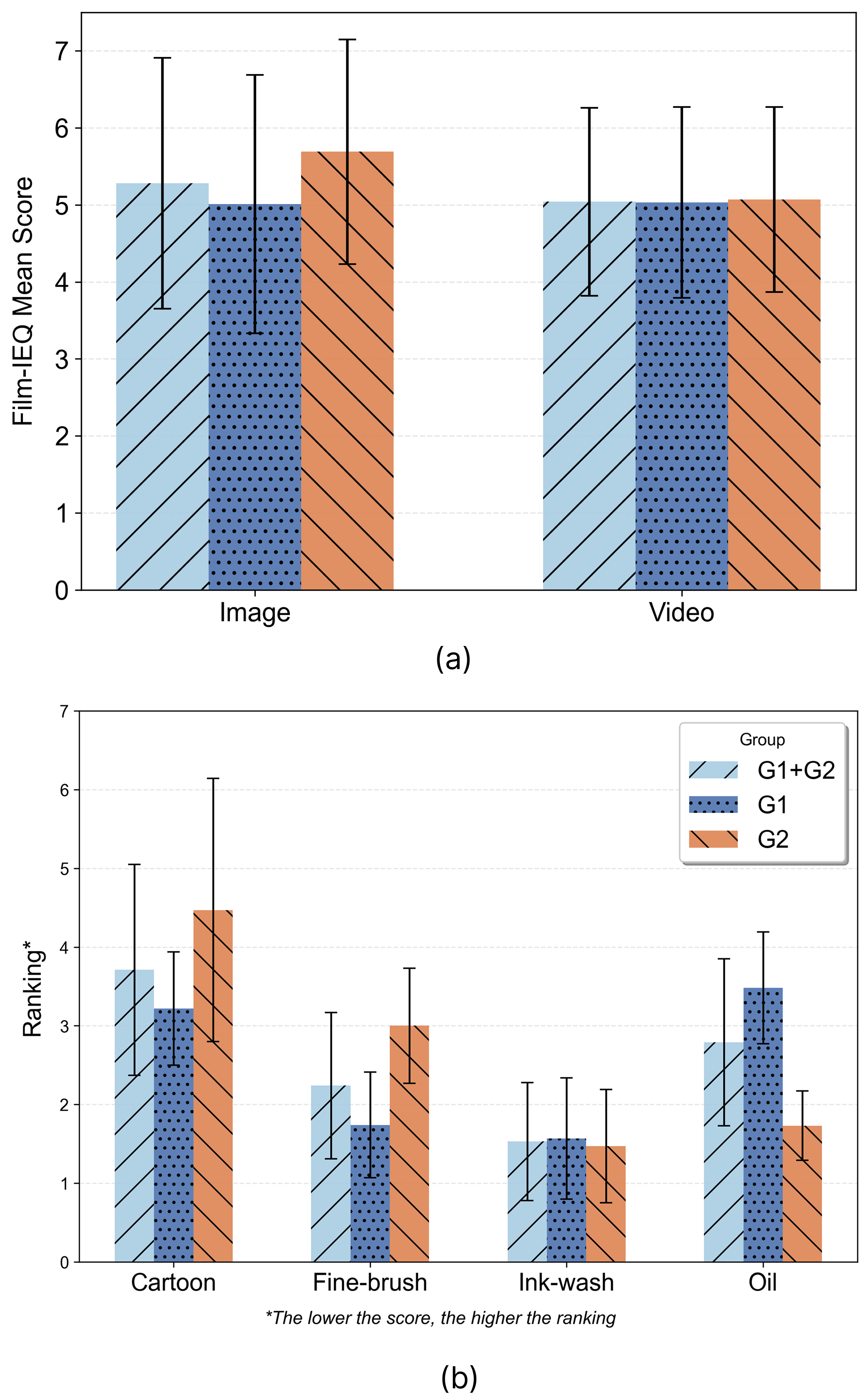}
    \hfill
    \caption{\textbf{Media format and visual style effects on Film-IEQ.} (a) Film-IEQ comparison of different modalities (image vs. video) in the AI symbolic path (G1+G2). Static key frames are rated slightly higher in interpretability, engagement, and symbol readability. (b) Style ranking in G1+G2, showing ink-wash/fine-brush preferred over oil/cartoon (including G1 and G2 separately).}
    \Description{A vertically arranged figure with two charts illustrating media format and visual style effects on Film-IEQ. The top chart (a) compares Film-IEQ scores between image-based key frames and video clips in the AI symbolic path for groups G1 and G2, showing that static images are rated slightly higher in interpretability, engagement, and symbol readability. The bottom chart (b) presents a ranking of visual styles in G1 and G2, indicating that ink-wash and fine-brush styles are preferred over oil and cartoon styles for supporting understanding.}
    \label{fig:7}
\end{figure}

\textbf{Differences between images and videos.}
To compare the effectiveness of two media formats (images and video) in helping users understand core cultural activities, we performed a paired-sample $t$ signed-rank test on the Film-IEQ scores in both image and video conditions. The Film-IEQ scale assesses how well users grasp the activities, intentions, and key symbols conveyed by the media content. Media format matters for interpretability. In G1+G2 (Figure~\ref{fig:7}a), Film-IEQ scores in the image condition were $M = 5.28$ ($SD = 1.01$), and in the video condition $M = 5.04$ ($SD = 0.74$). The difference was significant, $t = 2.18, p = .036$. Participants generally felt that static key frames were slightly better than short clips at clarifying ``who is doing what'' and ``what is going on.''

Qualitatively, participants in G1 first built a structure from multiple images and then used video to complement motion and atmosphere. Participants in G2, after embodied activities, treated images as a way to ``review and organize'' the structure, and video more as emotional and dynamic cues. This suggests that within the AI symbolic path, images are better suited to clarifying structure, while video mainly contributes dynamic and affective information; the two formats are functionally complementary for the multimodal understanding targeted in RQ1.

\textbf{Effects of visual style and semantic consistency.}
We further summarized the ranking of four styles of AIGC in terms of how much they helped understanding. As shown in Figure~\ref{fig:7}b, ink-wash ($M = 1.53$, $SD = 0.76$) and fine-brush ($M = 2.24$, $SD = 0.94$) outperformed oil ($M = 2.79$, $SD = 1.07$) and cartoon ($M = 3.71$, $SD = 1.35$), with lower values indicating better ranking. There was no significant interaction between style and order.

This ranking aligns with feedback from the interviews about ``cultural authenticity and brushwork lexicon matching'': when the visual style matches the cultural symbols it carries, participants more easily identify symbols and associate them with relevant contexts. In contrast, cross-cultural styles (e.g., oil's texture and skin tone) and modern cartoon abstractions weaken semantic fit, increasing cognitive load. For design, this implies that when selecting styles of AIGC, priority should be given to optimizing style-semantic alignment, with ink-wash/fine-brush as primary narrative carriers, and oil/cartoon used for contrast or embellishment, to avoid disrupting the clarity of cultural cues.

\subsubsection{Cultural Resonance Score (CR) and ITC-SOPI: A Quantitative Trajectory from ``Perceiving'' to ``Resonating''}\label{sec6.1.2} 
\begin{figure}[htbp]
    \centering\includegraphics[width=0.37\textwidth]{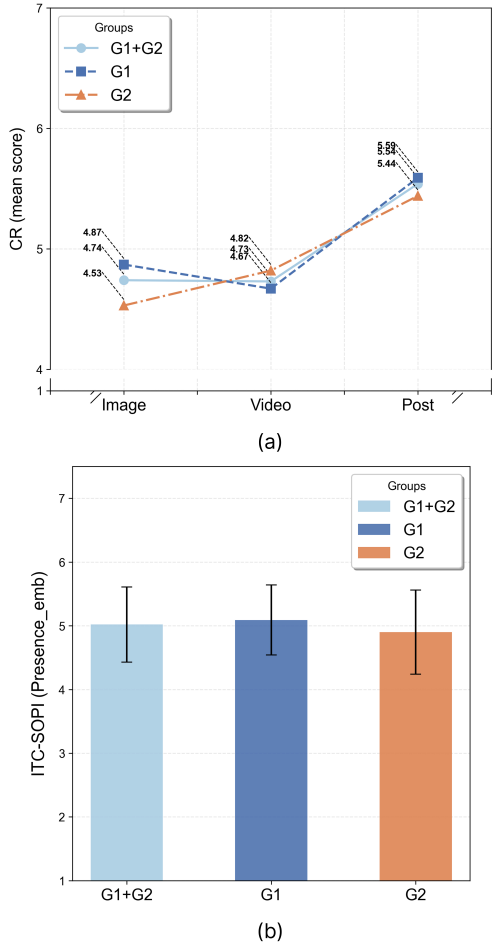}
    \caption{\textbf{Cultural resonance trajectory and embodied presence.} (a) CR trajectories at three stages ((after) image, (after) video, post (study)) for G1+G2, G1, G2 (95\% CI). (b) Presence\_emb (ITC-SOPI mean) for G1+G2, G1, G2 (95\% CI). Presence\_emb and CR\_post are positively correlated in the full sample ($\rho=.66, p<.001$).}
    \Description{A vertically arranged figure with two charts illustrating cultural resonance and embodied presence. The top chart (a) shows cultural resonance scores at three stages—after image viewing, after video viewing, and after the full study—for groups G1 and G2, indicating an increase in resonance after the embodied stage. The bottom chart (b) presents embodied presence scores measured by ITC-SOPI for G1 and G2, and shows a positive relationship between embodied presence and post-study cultural resonance. Together, the panels depict a quantitative progression from perceiving AI-mediated content to resonating through embodied experience.}
    \label{fig:8}
\end{figure}

We then examined the relationship between cultural resonance and embodied presence (Presence\_emb). CR was computed as the mean of three items on emotion, understanding, and aesthetics (Figure~\ref{fig:8}a). For G1+G2, CR\_img at the image stage was $M = 4.74$ ($SD = 0.94$), and CR\_vid at the video stage $M = 4.73$ ($SD = 1.08$); the difference was not significant, $t = 0.06$, $p = .95$. After the whole experience, CR\_post rose to $M = 5.54$ ($SD = 0.60$), significantly higher than the video stage, $t = -4.90$, $p < .001$. This indicates that textual guidance and the AI symbolic path had already raised CR to a medium-high level, but the major increase occurred after the embodied path and overall integration.

Embodied presence was represented by Presence\_emb, the mean ITC-SOPI score for the embodied stage (Figure~\ref{fig:8}b). For G1+G2, Presence\_emb was $M = 5.02$ ($SD = 0.59$); G1 and G2 scored $M = 5.09$ ($SD = 0.55$) and $M = 4.90$ ($SD = 0.66$) respectively, with no significant difference ($p = .37$). Presence\_emb was strongly correlated with CR\_post, $\rho = .66$, $p < .001$, $N = 23$, while prior familiarity with literati gatherings was not reliably correlated with CR\_post ($\rho = -.39$, $p = .067$) or Presence\_emb ($\rho = -.16$, $p = .46$). Even participants who were initially unfamiliar with literati gatherings could reach higher cultural resonance if they experienced sufficient presence in the embodied stage.

In summary, text and AI images/video mainly ``clarify what is happening,'' providing a stable semantic and emotional starting point, whereas the embodied stage deepens cultural meaning on top of a strong sense of ``being there.'' Quantitatively, this supports the dual-path division of labor: AI provides symbolic accessibility; embodiment provides experiential presence.

\begin{figure}[h]
    \centering
    \includegraphics[width=0.37\textwidth]{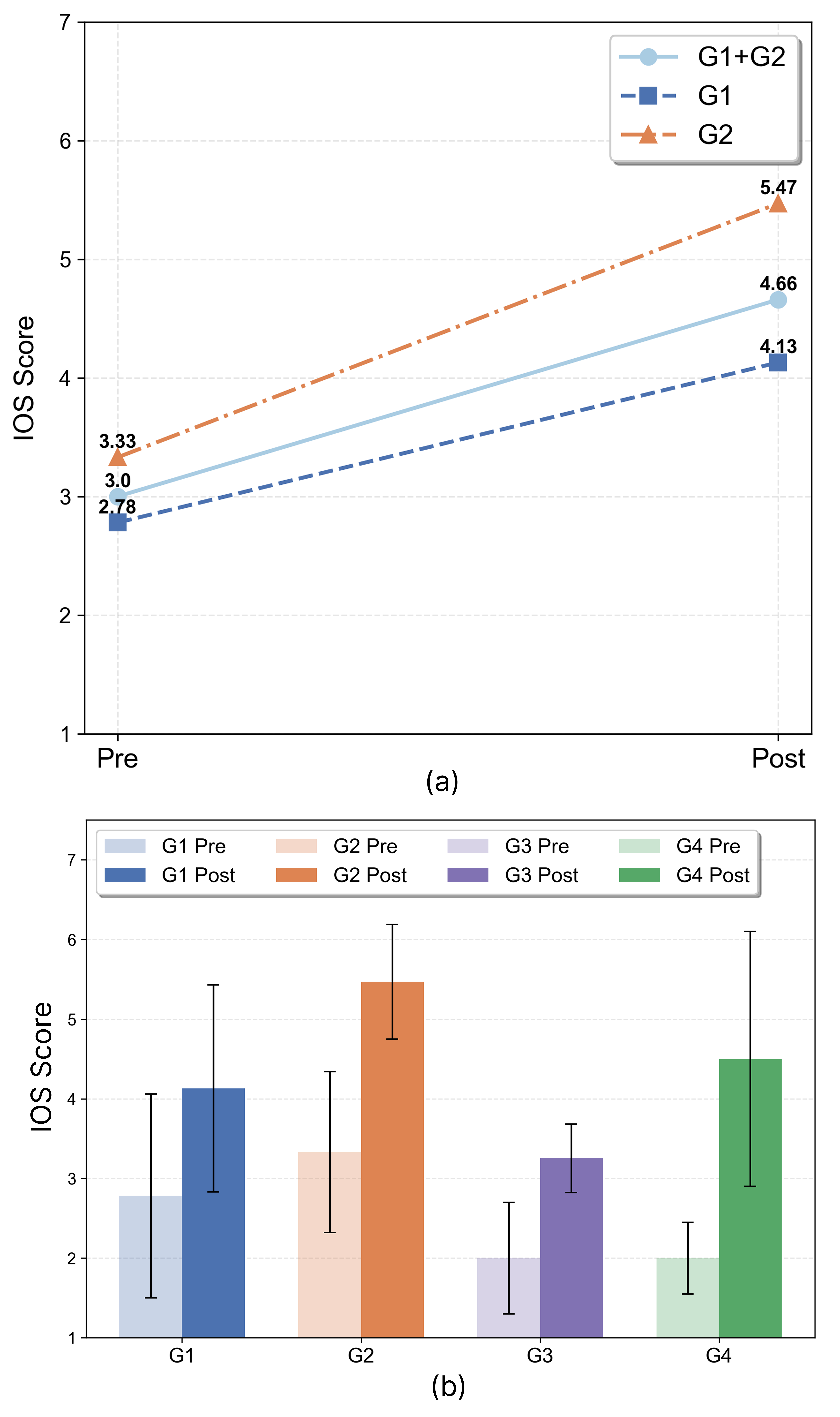}
    \caption{\textbf{Changes in psychological closeness.} (a) Pre (study) vs. post (study) IOS scores for G1+G2 (from 3.00 to 4.66, t(37) = 9.53, p<.001). (b) Pre- vs. post IOS means for all participants (visualized with tone contrast rather than bars).}
    \Description{A vertically arranged figure illustrating changes in psychological closeness measured by the Inclusion of Other in the Self (IOS) scale. The top panel (a) shows IOS scores increasing from pre-test to post-test for the combined groups G1 and G2, indicating greater psychological closeness to literati gatherings after the study. The bottom panel (b) compares IOS changes across two order conditions, showing larger gains when the embodied experience precedes the AI-media phase than when the order is reversed. Together, the figure highlights overall improvement in closeness and an order effect between embodied and AI-mediated experiences.}
    \label{fig:9}
\end{figure}

\begin{table*}[htbp]
\centering
\footnotesize
\caption{\textbf{KANO model classification results for system feature requirements.} This table displays the KANO classification results for 10 system features. The categories A (Attractive), O (One-dimensional), M (Must-be), I (Indifferent), R (Reverse), and Q (Questionable) are determined based on user survey responses. $CS^+$ (Better Index) and $CS^-$ (Worse Index) indicate the satisfaction and dissatisfaction coefficients, respectively.}
\label{tab:kano_full_compact}
\begin{tabular}{cccccccccccc}
\toprule
\multirow{2}{*}{ID} & \multirow{2}{*}{Feature} & \multicolumn{6}{c}{KANO Categories} & \multirow{2}{*}{\makecell[c]{KANO\\Classification}} & \multirow{2}{*}{\makecell[c]{Better Index\\(CS+)}} & \multirow{2}{*}{\makecell[c]{Worse Index\\(CS-)}} \\
\cmidrule(lr){3-8}
 & & \textbf{A} & \textbf{O} & \textbf{M} & \textbf{I} & \textbf{R} & \textbf{Q} & & & \\
\midrule
F1 & Content Breadth        & 40.48\% & 11.90\% & 2.38\%  & 45.24\% & 0.00\% & 0.00\% & Indifferent & 52.38\% & -14.29\% \\
F2 & Social Interaction     & 52.38\% & 21.43\% & 0.00\%  & 26.19\% & 0.00\% & 0.00\% & Attractive & 73.81\% & -21.43\% \\
F3 & Environmental Immersion & 59.52\% & 16.67\% & 0.00\%  & 23.81\% & 0.00\% & 0.00\% & Attractive & 76.19\% & -16.67\% \\
F4 & Ritual Participation   & 59.52\% & 16.67\% & 0.00\%  & 21.43\% & 2.38\% & 0.00\% & Attractive & 78.05\% & -17.07\% \\
F5 & Exploration Freedom    & 54.76\% & 21.43\% & 0.00\%  & 23.81\% & 0.00\% & 0.00\% & Attractive & 76.19\% & -21.43\% \\
F6 & Personalization        & 59.52\% & 19.05\% & 0.00\%  & 21.43\% & 0.00\% & 0.00\% & Attractive & 78.57\% & -19.05\% \\
F7 & Art Appreciation       & 57.14\% & 9.52\%  & 2.38\%  & 30.95\% & 0.00\% & 0.00\% & Attractive & 66.67\% & -11.90\% \\
F8 & Multiplayer Social     & 45.24\% & 16.67\% & 0.00\%  & 38.10\% & 0.00\% & 0.00\% & Attractive & 61.90\% & -16.67\% \\
F9 & Knowledge Guidance     & 23.81\% & 9.52\%  & 0.00\%  & 61.90\% & 2.38\% & 2.38\% & Indifferent & 35.00\% & -10.00\% \\
F10 & AI Real-time Feedback & 28.57\% & 11.90\% & 0.00\%  & 59.52\% & 0.00\% & 0.00\% & Indifferent & 40.48\% & -11.90\% \\
\bottomrule
\end{tabular}
\end{table*}

\subsubsection{IOS: Psychological Closeness and Order Effects}\label{sec6.1.3}
The Inclusion of Other in the Self (IOS) scale measures participants' psychological closeness to ``literati gatherings'' as a cultural object, and is a key indicator for whether the dual-path truly ``reduces distance'' between people and culture. In G1+G2 overall (Figure~\ref{fig:9}a), IOS increased from $M = 3.00$ ($SD = 1.23$) at pre-test to $M = 4.66$ ($SD = 1.30$) at post-test, with a significant difference, $t = 9.53$, $p < .001$. Considering order reveals more detail: in G1, IOS rose from $M = 2.78$ to $M = 4.13$ ($\Delta\text{IOS} \approx 1.35$); in G2, from $M = 3.33$ to $M = 5.47$ ($\Delta\text{IOS} \approx 2.13$). The difference in gain was significant, $t = -2.25$, $p = .033$, and the Phase $\times$ Order interaction was also significant in the mixed model.

These results show that even in a short session, psychological closeness improves substantially whether participants encounter AI or embodiment first. This suggests that such experiences are not merely informational or entertaining, but can strengthen users’ psychological connection to cultural practices. The larger gain in the ``Embodied to AI'' sequence further suggests that embodied experience has a stronger initiating effect on reducing distance, while subsequent AI content in that sequence mainly consolidates and structures the experience. This order effect is consistent with interview and observational findings in Section~\ref{sec6.2}.

\subsubsection{SAM: A Stable Emotional Background for the Understanding Process}\label{sec6.1.4}
For emotional experience, the three SAM dimensions generally remained in the medium to medium–high range across phases. In G1+G2, valence was $M = 4.16$ ($SD = 1.50$) at pre-test, $M = 4.68$ at the image stage, $M = 4.42$ at the video stage, and $M = 3.29$ at post-test. Arousal fluctuated between 4.89 and 5.24, and dominance stayed at moderate levels, with a slight decline only at the end.

This matches our design goal: the system does not aim for emotional ``peaks,'' but for a stable, comfortable emotional background in which participants can move from symbol recognition to internalized meaning. Together with CR, Presence\_emb, and IOS, SAM can be understood as an emotional ``floor'': emotion motivates engagement, while major changes occur in structural understanding, presence, and identification rather than in raw arousal.

\subsubsection{KANO: Functional Elements and Satisfaction Structure}\label{sec6.1.5}
To further address RQ3 ``Which design elements truly drive the experience, and which are more like optional extras?'' we conducted KANO analysis on the collected questionnaires, computing Attractive/ One-dimensional/ Must-be/ Indifferent/ Reverse/ Questionable (A/O/M/I/R/Q) categories and $CS^+$/$CS^-$ coefficients for 10 features (F1--F10), as shown in Table~\ref{tab:kano_full_compact}.

The results show a clear pattern: seven features (F2--F8) are mainly Attractive qualities, while three features (F1, F9, F10) are Indifferent; none are dominated by Must-be or One-dimensional categories (Table~\ref{tab:kano_full_compact}). Participants did not treat any single feature as ``basic configuration,'' but saw the system as a bundle of cultural experience features where ``doing it well'' brings noticeable gains.

\begin{figure}[h]
    \centering
    \includegraphics[width=0.9\linewidth]{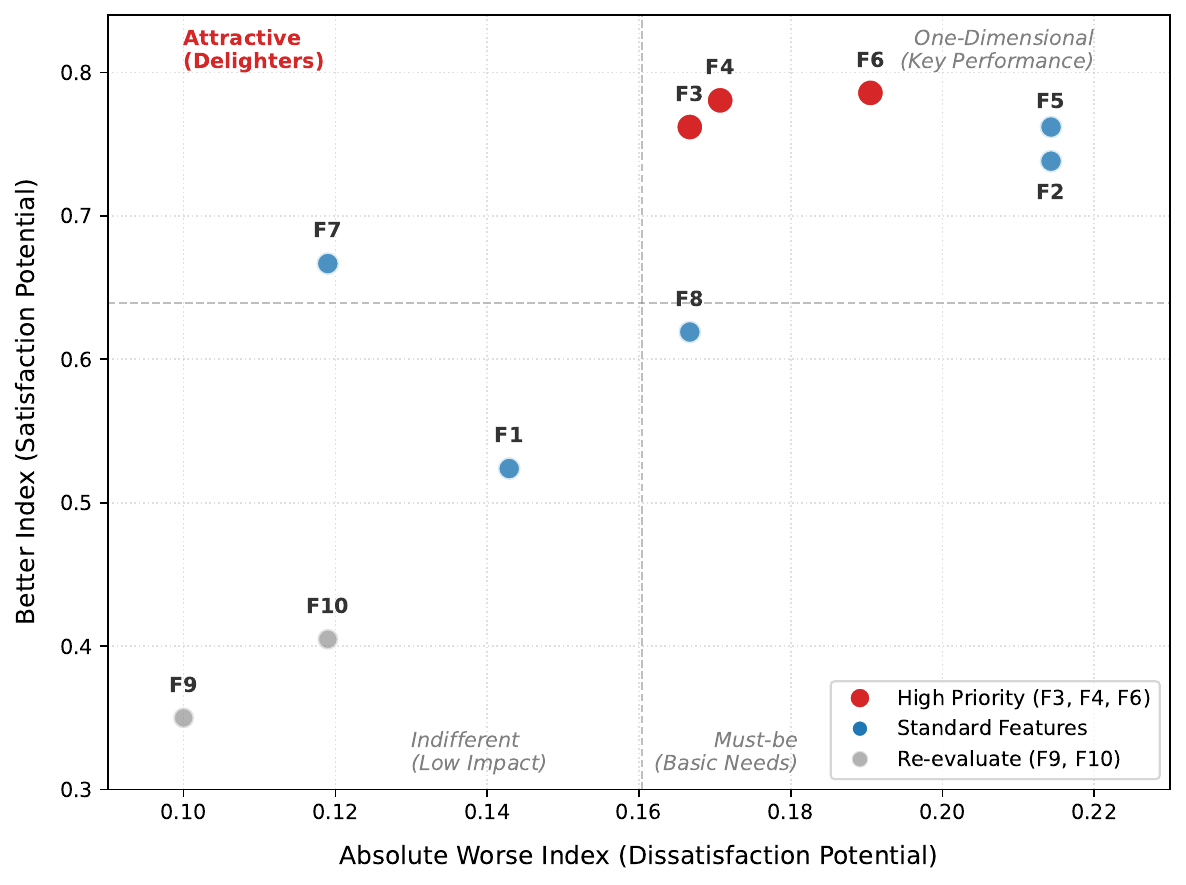}
    \caption{\textbf{KANO overview in Better-Worse quadrant.}}
    \Description{A scatterplot showing a KANO overview in the Better–Worse quadrant, with $CS^+$ (better) on the horizontal axis and $CS^-$ (worse) on the vertical axis, including overall mean reference lines. Scatterplot of $CS^+$ (better) and $CS^-$ (worse) with overall means. Seven features (F2–-F8) fall in the Attractive region (F6 with highest $CS^+$, 78.57\%; F5 with largest $|CS^-|$, -21.43\%), while F1/F9/F10 cluster near the Indifferent region ($CS^+$ = 35.00\% for F9, 40.48\% for F10).The plot visualizes how different design features contribute to user satisfaction and dissatisfaction within the system.}
\label{fig:KANO}
\end{figure}

Satisfaction sensitivity shows priorities: personalization (F6) has the highest $CS^+$ (78.57\%), followed by ritual participation (F4, 78.05\%), with environmental immersion (F3) and exploration freedom (F5) also high (both 76.19\%). These features markedly increase satisfaction when present. Social interaction (F2) and exploration freedom (F5) have the largest absolute $CS^-$ (both $-$21.43\%), indicating that their absence most easily causes dissatisfaction. In contrast, content breadth (F1), knowledge guidance (F9), and AI real-time feedback (F10) are often treated as ``nice to have'': they exist functionally, but with lower $CS^+$ and near-zero $CS^-$ (e.g., 35.00\% for F9 and 40.48\% for F10) (Figure~\ref{fig:KANO}).

Overall, the KANO results align with previous quantitative patterns: within the AI-driven dual-path framework, what truly shapes experience quality is not ``how visible the AI is'' or ``how complete the knowledge explanation is,'' but whether participants are invited into ritual participation, given social structures and spaces to explore, and allowed to ``be one of the gathering.'' These are core design levers at the intersection of the embodied experience path and the AI symbolic path.

\begin{figure*}[ht]
   \centering
   \includegraphics[width=0.7\textwidth]{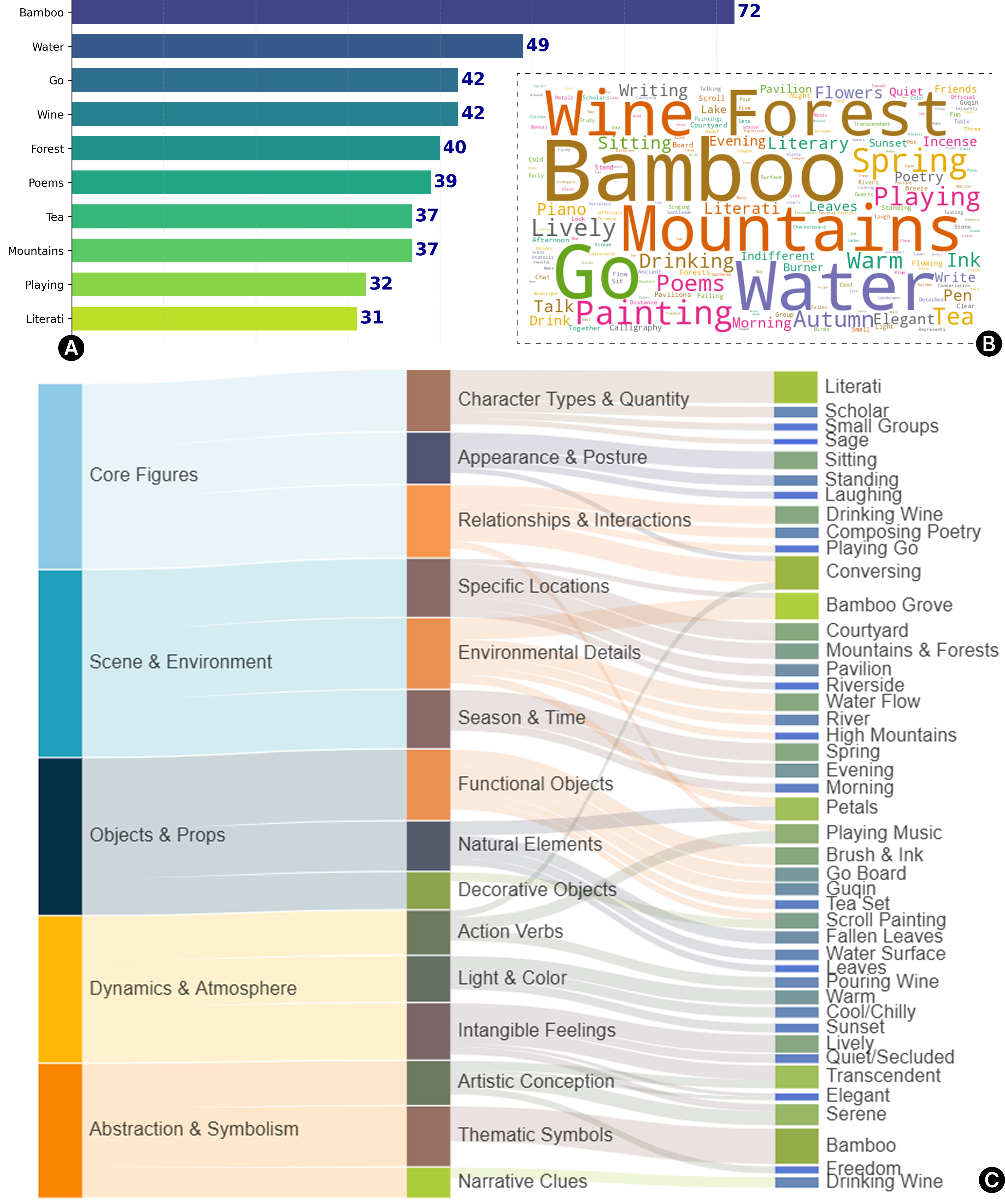}
   \caption{\textbf{Visual analysis of thematic and lexical patterns in user-generated association keywords.} (A) Terms with a frequency of 31 or higher, featuring prominent keywords such as ``Bamboo,'' ``Water,'' and ``Go'' that reflect the focus on artistic and natural motifs. (B) The word cloud of high-frequency terms. These terms primarily belong to three conceptual clusters: ``Literary Elements,'' ``Natural Terms,'' and ``Cultural Activity,'' providing quantitative insights into the corpus's thematic composition. (C) Sankey diagram between 15 descriptive categories and 5 overarching thematic groups, demonstrating how textual elements contribute to broader thematic constructs.}
   \Description{A multi-part visualization analyzing user-generated association keywords for ``Wenren Yaji'' from groups G1 and G2. Panel (a) shows terms with a frequency of 31 or higher, highlighting prominent keywords such as ``bamboo,'' ``water,'' and ``Go,'' which emphasize artistic and natural motifs. Panel (b) presents a word cloud of high-frequency terms, clustered into three main conceptual groups: literary elements, natural terms, and cultural activities. Panel (c) displays a Sankey diagram linking 15 descriptive categories to five overarching semantic dimensions—environment, material media, social relations, actions, and affective–aesthetic tone—illustrating how individual keywords contribute to broader thematic structures.}
\label{fig:association_analysis}
\end{figure*}

\subsection{Qualitative Findings}\label{sec6.2}
\subsubsection{Association Transfer and User Interpretation Paths}\label{sec6.2.1}

In G1 and G2, participants wrote associative words about \textit{``Wenren Yaji''} before the experience (Figure~\ref{fig:association_analysis}). Together, they produced 3,214 word occurrences (G1: 2,104; G2: 1,110), corresponding to 706 and 462 unique word forms. High-frequency words were manually grouped into five semantic dimensions: environment (nature and setting), material media (Go, tea, wine, writing tools, etc.), social relations (roles and groups), actions (behavior and posture), and affective-aesthetic tone (atmosphere and subjective feeling). We then focused on core words in these five dimensions (G1: 1,068; G2: 555; total: 1,623).

Overall, both groups showed similar ``symbolic foundations'': participants first constructed a static scene around ``where it is'' and ``what is there.'' Environmental words accounted for about 30\% and material media about 27\%; high-frequency terms concentrated on natural imagery (bamboo, water, forest, mountains) and objects (Go, tea, wine). This echoes the quantitative finding that the AI symbolic path ``first clarifies the scene,'' while also showing that this level is mainly about location and objects.

More important changes appeared in social, emotional, and action dimensions. Many participants used literati and scholars instead of general terms like people or friends, indicating a shift toward seeing specific cultural identities. Emotional terms moved from broad lively, warm to more nuanced, refined, calm, secluded, and relaxed. Action words included not only play and drinking but also composing, chanting, wielding, and pouring, which directly match activity scripts such as composing poetry, chanting, writing, and pitch-pot.

In summary, the AI symbolic path helps participants grasp the visible scene and objects, while the embodied experience path weaves these elements into a lived script with roles, emotions, and repeatable actions. This transition is difficult to capture with quantitative measures alone, but the associative words provide independent linguistic evidence for the dual-path mechanism proposed in RQ3.

\subsubsection{Complementarity Between AI Symbolic and Embodied Paths}\label{sec6.2.2}

Based on semi-structured interviews with G1 (P1--P23) and G2 (P24--P38), we supplement the quantitative findings by examining: how AI-generated content serves as an entry point, how embodied experience builds presence, and how the two paths work together.

\textbf{AI-generated content as an entry point and ``problem space.''}
Many participants described AI images and videos as their first step into the literati gathering context. P6 said that with text alone, it is ``hard to imagine what exactly happens in a gathering just from a few words.'' P3 felt ink-wash and fine-brush styles were ``the most elegant,'' P27 found them consistent with prior impressions of ancient people, and P34 thought they ``make these ancient elements move'' and make the background ``more immersive,'' more engaging than static panels. In contrast, P14 described oil painting as ``too Western'' and cartoons as ``too modern,'' which broke immersion. These evaluations match the style rankings in 6.1.1 and show that style–context fit is crucial for AI content to function as an understanding entry point (RQ1).
At the same time, participants were very sensitive to AIGC issues in physical logic and detail. P2 and P10 noted that some pitch-pot and instrument-playing motions ``did not match real-world logic.'' P26 called the AI videos ``the strangest part,'' with ``many weird places that pull you out,'' and P27 pointed out clipping and hand positions that drew attention away from content. P9 summarized: ``It gives me a rough feeling; I treat the details as AI bugs.'' After embodied experience, P31 felt that ``before I didn’t really understand pitch-pot; after doing it and then seeing the video, even if the picture is not quite right, I can still guess what it wants to express.''
Overall, the AI symbolic path quickly makes the scene understandable when style and composition are appropriate, while its visible flaws expose gaps between current AI and cultural nuance, leaving room for the embodied path to repair and enrich meaning.

\textbf{The embodied experience path builds a sense of presence through social interaction and rituals.}
Participants also recognized ritual and li. P24 reflected that mutual politeness in Go made him feel that it was ``not just casual play, but a kind of \textit{Li}.'' P7 described the poetry segment as ``a ritual rather than a casual game.'' P8 and P9 called pitch-pot a ``social catalyst'': when someone scores, ``everyone is happy,'' and even misses cause laughter, a ``social attribute'' that gives extra meaning.
The environment amplified this presence. P33 felt it was different from ``playing at the Go club,'' because there was a specific literati setting and music: ``doing the same activity in such a setting makes it easier to feel that I am participating in a gathering.'' P34 thought that ``even just Gomoku, with this background music, feels better, more immersive.'' P16 and P23 mentioned that lights, \textit{guqin} music, and incense made people ``lower their voices as soon as they enter''; P35 felt that ``scene and music promote immersion.''
At the same time, participants recognized limits. P33 admitted being ``more interested in the activities'' and tending to ignore the deeper background. P38 felt that, as participants, they still held ``a playful, appreciative attitude'' and did not fully enter the historical context, but the activities still ``brought me closer to ancient literati gatherings.'' This matches ITC-SOPI and IOS patterns in Section~\ref{sec6.1}: the embodied path reliably provides presence and closeness, but deeper cultural understanding depends on scripting and guidance, not ``fun'' alone (RQ2).

\textbf{Synergy between the two paths.} 
Interviews overall show a clear division of labor and synergy between AI, symbolic, and embodied paths. Embodied experience was widely seen as the main channel for reducing psychological distance. P30 described pitch-pot and Go as ``the best parts,'' because ``you can actually participate,'' and playing with others makes you ``naturally feel more related to them.'' P32 said that after doing the activities with friends, it was easier to imagine ``ancient people also gathering like this,'' turning literati gatherings from a distant historical scene into something one could be in.
Correspondingly, the AI symbolic path mainly builds the framework and imagery. For participants unfamiliar with the activities, P20 felt that one ``does need videos and pictures first to know what to do; then it is easier to get into it.'' For participants familiar with poetry, P18 said that AI-generated images ``help draw out the picture in my head,'' which is then ``enacted once'' through embodiment. P38 concluded that for those ``not steeped in this culture,'' using only one path ``is worse than using both.''
In short, participants generally treated the AI symbolic path as a structured ``see and understand'' entry, and the embodied path as the core channel for ``being in it.'' The former tells them ``what a literati gathering looks like,'' while the latter lets them feel ``I could also do this,'' jointly moving understanding from scene recognition to situational identification.

\subsubsection{Exploratory Findings from Cross-Cultural Participants and Children}\label{sec6.2.3}
This subsection focuses on G3 (non-Chinese cultural background individuals, N = 4) and G4 (children, N = 6), using descriptive trends and interview excerpts to exploratorily examine the dual-path framework in cross-cultural and developmental contexts. No inferential statistics are reported. Quantitatively, IOS scores in both groups tended to increase over time (Figure~\ref{fig:9}b), suggesting that the dual-path can also work in heterogeneous samples; differences lie mainly in how they use the two paths.

\textbf{Cross-cultural participants: text/AI as ``manual,'' activities as ``core.''}
For G3, text and AI content were experienced as a ``necessary but effortful manual,'' while embodied activities formed the experiential core. One participant likened reading classical text to ``high school reading comprehension,'' hoping for ``audio text'' and more explanation; another felt that presenting ancient texts only on paper ``wastes the equipment'' and preferred integration with on-screen media. They generally reported that ``actually doing something is easier to understand'' (P40, P42).

\textbf{Children: embodiment as ``universal language,'' AI as ``funny background.''}
G4 more strongly highlighted the role of embodiment. Children laughed while watching AI-generated videos, pointing out ``bugs,'' such as ``why is the water floating, the force is reversed,'' ``this is called an experiment? It’s just watching videos,'' and ``AI is not ready yet, it’s very unsafe'' (P43––P48), treating AI more as ``funny videos'' than historical representation. They felt that the virtual content in the videos was less appealing than actually throwing, playing, or writing.

Overall, comparing G3 and G4 with G1+G2 suggests that the dual-path framework has some generality across groups, but the relative weight of AI symbolic vs. embodied paths should be adjusted to educational level and cultural background. For cross-cultural participants, the explanatory function of text and AI should be strengthened and complemented by embodied activities. For children, embodied experience should be treated as the main language, with AI content positioned as auxiliary cues and triggers for discussion.
\section{Discussion}\label{discussion}

This study provides initial empirical support for our AI-driven dual-path framework for cultural understanding through a mixed-methods design. Our main contribution lies in integrating two seemingly independent paths—the AI-driven symbolic path, which focuses on visual style and technical rendering to make cultural symbols readable and culturally ``natural'', and the embodied experience path, which relies on embodied tasks, ritualized interaction and multisensory cues to build social presence, ritual sense and immersion—into a single, coherent process of cultural understanding. Across Section~\ref{results}, the quantitative and qualitative results together indicate that these two paths support different stages of seeing, perceiving, and resonating, and that their combination can lead to a stronger cultural connection than either path alone.

This dual-path approach emphasizes that merely pursuing the ``naturalness'' of visuals is insufficient to ensure deep cultural understanding. Cultural depth can only be effectively transferred when technology can explicitly highlight core cultural symbols and guide the body into a structured ritual. This approach is not only applicable to literati gatherings but also offers insights for other complex cultural practices, highlighting the integration of multimodal content generation with embodied activity design, as well as the complementarity of symbol interpretation and bodily experience. In this process, AI technology is not meant to replace human cultural heritage but to serve as a powerful tool to enhance cultural understanding. Below, we will further elaborate on our findings in relation to previous theoretical and empirical work.

\subsection{AI-Generated Visual Content: A Bridge Full of Gaps}\label{sec7.1}

Quantitative results show that AI-generated content in the AI-driven symbolic path can substantially clarify what is happening in a complex cultural scene. In Section~\ref{sec6.1.1}, Film-IEQ scores for both images and videos were above the mid-point of the scale, and participants rated AI imagery as helpful for understanding ``who is doing what'' in the gathering. Cultural resonance during the image and video phases ($CR\_img$ and $CR\_vid$, around 4.7 on a 7-point scale) also reached a stable mid–high level before any embodied activity. These findings align with prior work on text-to-image/video generation for cultural visualization, which emphasizes its ability to lower the modeling threshold and quickly render scene-level structures.

However, when moving into dynamic, complex scenarios, the limitations of current AI technology become apparent, leading to cognitive dissonance similar to the ``uncanny valley'' effect~\cite{mori2012uncanny, seyama2007uncanny}. The negative emotional experiences triggered by AI-generated videos are directly related to user criticisms of the videos' lack of physical credibility, temporal coherence, and emotional expression. The interviews pointed to issues with micro-level credibility (stiff expressions, distorted actions, and sizing), aligning with the uncanny valley effect, where high realism but insufficient authenticity undermines immersion and likability. This phenomenon indicates that AI still has significant shortcomings in mastering physical credibility and subtle emotional expression, limiting its utility in cultural reproduction projects that require a high degree of realism and emotional resonance.

Style preferences further clarify how the AI-driven symbolic path should be positioned. As shown by the style ranking in Section~\ref{sec6.1.1} and echoed in interviews, we observed that users preferred ink-wash and fine-brush styles over more realistic ones, suggesting that in cultural contexts, styles that prioritize conceptualization and disclose uncertainty (informing users about potential physical/emotional discrepancies in AI-generated content) should be favored. AIGC should be positioned as a ``preliminary framework'' rather than a final representation. Meanwhile, the new generation of 3D/scene-level generation is laying the foundation for providing tools for cultural context construction \cite{spennemann2024generative}.

\subsection{The Central Role of Embodied Interaction: From Passive Observer to Active Participant}\label{sec7.2}

One of the most profound discoveries of this study, and its core contribution, is the critical role of embodied interaction in promoting cultural understanding and emotional connection. The data showing a significant improvement in IOS scores post-experience ($+1.66$) perfectly aligns with the repeated emphasis from participants in interviews on the importance of ``active participation'' (Section~\ref{sec6.1.3}). By introducing procedural social-ritual scripts (greetings, turn-taking, cup passing) and tactile/sonic elements into the ``human-object-field'' dynamic, both IOS and CR increased, consistent with the embodied cognition and in-situ practice perspectives \cite{suchman2007human}, where understanding originates from bodily presence and action. Compared to many systems focused on single artifacts or individual experiences \cite{zhao2025immersive}, we emphasize the explicit design of social presence and role transitions, quantifying the benefits with ITC-SOPI. Pre-test data showed that participants had a relatively distant psychological distance from the ``literati gathering'' concept. However, through personal engagement in activities such as pitch-pot and calligraphy, they not only understood the activities themselves but, more importantly, felt an emotional resonance, such as ``experiencing what the ancients might have felt'' (P7). From a design standpoint, embodied narratives aid in cross-cultural understanding, supporting our hypothesis of the ``move from presence to resonance'' path \cite{giannopoulos2008effect, lessiter2001cross}. This active, hands-on participation transforms cultural concepts from abstract knowledge into tangible, perceptible experiences, suggesting that the best positioning of technology in cultural experiences is not as a replacement for reality, but as a bridge to real-life experience.

\subsection{KANO-Based Design Implications for AI-Driven Dual-Path Experiences}\label{sec7.3}

The KANO analysis in Section~\ref{sec6.1.5} offers a complementary, user-centered view of which system elements matter most in an AI-driven dual-path design. Most features related to social structure, environment, ritual participation, exploration freedom, personalization, and aesthetic appreciation were classified as Attractive, with high positive satisfaction sensitivity ($CS^+$). In contrast, content breadth, explicit knowledge guidance, and visible AI real-time feedback were predominantly indifferent, with low $CS^+$ and near-zero $CS^-$\cite{hassenzahl2010experience}. Experimental results show that users' expectations for cultural experiences have moved beyond traditional, didactic ``educational'' models. They no longer wish to passively receive information but crave an experience that offers autonomy, creativity, and social connections. Future cultural heritage systems should shift their design focus from ``teaching knowledge'' to ``stimulating exploration and pleasure.''
Based on these findings, we propose a series of Design Implications  \cite{mattelmaki2005applying, gaver1999design, sanders2014probes} aimed at transforming user needs into testable, iterative design directions.

\begin{itemize}
 \item\textbf{Probe A: Physical Credibility of Visual Symbols.}  
By providing accurate and physically plausible visual cues, we can avoid cognitive dissonance or ``misreading,'' thus enhancing immersion and AI credibility.

 \item\textbf{Probe B: Visualization of Social Frameworks and Turn-Taking Mechanisms.}  
By visualizing abstract social rules, we reduce uncertainty in interactions, strengthen social presence, and guide users to engage in collective behaviors like ``singing along'' or ``co-creating.''

 \item\textbf{Probe C: Creation of Multisensory Atmosphere.}  
Going beyond visual and auditory experiences, activating the user's sensory system through multisensory input creates a richer, more layered sense of immersion and pleasure.

 \item\textbf{Probe D: Ritualization of Behavioral Flow.}  
By concretizing the abstract sense of cultural rituals into executable physical actions, we help users internalize cultural meaning through action, enabling the cognitive shift from ``viewing'' to ``internalization.''

 \item\textbf{Probe E: AI Content Explainability and Style Management.}  
By proactively providing background information and explanations, we manage user expectations and turn AI's limitations into an acceptable artistic or technological style, avoiding negative emotions triggered by the ``uncanny valley'' effect.
\end{itemize}

\section{Limitations and Future Work}\label{limitations and future work}
This study has several limitations that point to directions for future work. First, while the main sample (G1+G2) is statistically adequate for our analyses, the cross-cultural group and the child group are relatively small and are used only for exploratory trend observation. Future research should include more diverse age ranges and cultural backgrounds to examine how generalizable and differentiated the AI-driven symbolic path and the embodied path are across populations. Moreover, although we counterbalanced the order of the symbolic and embodied paths across two groups, the per-group sample size was modest. As a result, our analysis may have lacked power to detect subtle order effects, and the current null finding on path order should be interpreted with caution.

Second, the embodied setting in this work was implemented as an indoor prototype installation. Although we incorporated physical artifacts and music, the historical context was not fully reconstructed. Follow-up work could build more gamified VR prototypes that embed pitch-pot, Go, and other activities into complete cultural narratives (e.g., role-based, script-driven experiences), and more systematically integrate spatial audio, scent, and other multisensory cues to enhance environmental immersion.

Third, the AI content used here reflects both the current capabilities and limitations of general-purpose generative models, including noticeable deviations in physical logic. Future work could construct more targeted cultural-activity datasets or customized generation pipelines to improve the physical plausibility and temporal coherence of dynamic AIGC, and deploy these design probes in real cultural settings such as museums or schools, iteratively refining social scripts, AI presentation styles, and embodied task design in situated use.
\section{Conclusion}\label{conclusion}
In this study, we proposed an AI-driven dual-path framework for cultural understanding, grounded in embodied cognition theory. We then instantiated this framework through \textit{GatheringSense}, an AI-driven dual-path cultural experience centered on Chinese literati gatherings (\textit{Wenren Yaji}), to examine how a cognitive relay between visual interpretation and embodied participation can be established.

Our findings consistently show that, while AI-generated visual media can serve as an effective front-end tool by providing a high-level conceptual frame, embodied, social, and multisensory interaction is crucial for a deeper cultural connection. Quantitative results show that embodied interaction significantly increases participants’ psychological closeness to the cultural concept, and we did not observe statistically significant differences between the two path orders on our learning-related measures. Qualitative analyses further reveal current limitations of AI in physical plausibility and emotional expression. Exploratory results from cross-cultural participants and children indicate that, by adjusting the relative weight and sequencing of the two paths according to textual literacy and cultural background, the same dual-path mechanism can be transferred to different user groups.

The KANO analysis offers a clear roadmap for future design, highlighting that users value ``attractive'' features that bring enjoyment, agency, and social connection more than didactic, content-heavy elements. Building on these findings, we derived five design implications that provide concrete methods for creating AI-driven dual-path experiences in digital cultural-heritage contexts.

Ultimately, the key conclusion of this study is that the digital representation of cultural heritage should not stop at ``seeing,'' but should boldly transition towards ``resonating.'' Future digital humanities systems should go beyond passive screens and create experiences that seamlessly integrate carefully designed, culturally authentic visual content with rich, hands-on, and socially-driven interactions. This will not only transform users from passive observers into active participants in cultural narratives but also breathe new life into the transmission of cultural heritage.

\begin{acks} 
This work was partially supported by the National Natural Science Foundation of China (No. U25A20384) and the National Social Science Foundation of China (No. 24VWB020). We thank all participants for their time and participation in this study and are grateful to Professor Junjie Zhang and Xingbei Chen from the Hong Kong University of Science and Technology (Guangzhou) for their valuable support and assistance throughout the project. Generative AI tools were used to support selected parts of the research process, such as prototyping and image and video generation. The authors assume full responsibility for the content, interpretation, and use of all AI-generated materials presented in this paper.
\end{acks}

\bibliographystyle{ACM-Reference-Format}
\bibliography{sample-base}

\onecolumn
\appendix
\section{Appendix}\label{appendix}

\subsection{Survey Questions}\label{A.1}

\subsubsection{Background Information}\label{A.1.1}

\begin{enumerate}
    \item \textbf{Age range?} \\
    Under 18; 18–30; 31–40; 41–50; 51+

    \item \textbf{Gender?} \\
    Male; Female; Other

    \item \textbf{Highest education?} \\
    Primary School; Middle School; High School; Bachelor; Master; Ph.D./Doctorate; Other

    \item \textbf{Native language?} \\
    Chinese; English; Other

    \item \textbf{Research area/discipline?} \\
    Open text

    \item \textbf{Exposure to classical Chinese culture (poetry, calligraphy, exhibitions, etc.)} \\
    7-point frequency: Never; Rarely; Occasionally; Sometimes; Often; Almost daily; Very frequently

    \item \textbf{Familiarity with literati gathering (\textit{Wenren Yaji})} \\
    7-point familiarity: Not at all; Hardly familiar; A little familiar; Aware but not in depth; Quite familiar; Very familiar; Expert/research background

    \item \textbf{Exposure to AI-generated images/videos} \\
    7-point frequency: Never; Rarely; Occasionally; Sometimes; Often; Almost daily; Very frequently

    \item \textbf{Experience with immersive media (VR/AR/XR)} \\
    7-point experience: None; Almost none; A little; Some; Quite experienced; Very experienced; Highly experienced
\end{enumerate}

\begin{figure}[h]
    \centering
    \includegraphics[width=0.7\linewidth]{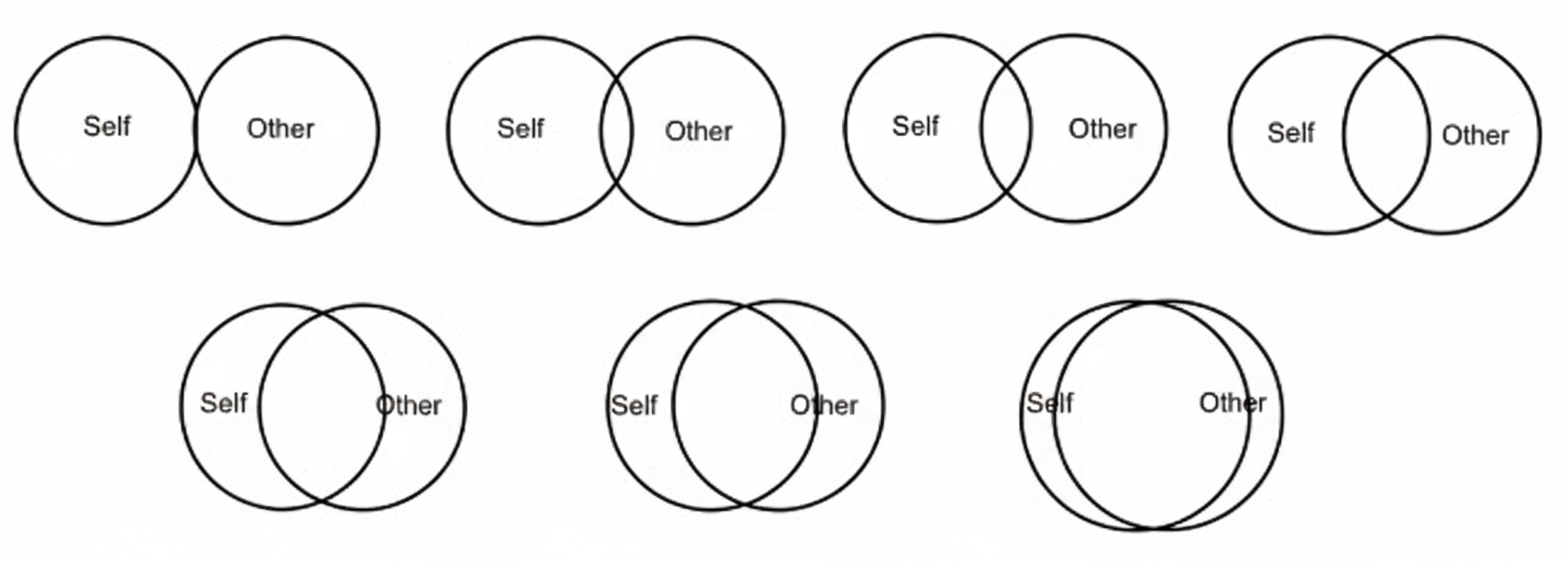}
    \caption{Inclusion of self with the literati gathering (\textit{Wenren Yaji}).}
    \Description{An illustration of the Inclusion of Other in the Self (IOS) scale using pairs of circles. The figure shows two circles representing the participant and the literati gathering (Wenren Yaji), with seven levels of overlap ranging from completely separate to almost fully overlapping. A smaller overlap indicates greater psychological distance, whereas a larger overlap indicates greater psychological closeness. The same visual scale is used for both pre-test and post-test measurements.}
    \label{fig:Appendix1}
\end{figure}

\subsubsection{The Inclusion of Other in the Self Scale (IOS) (Figure~\ref{fig:Appendix1})}\label{A.1.2}

\begin{itemize}
    \item \textbf{IOS (pre).} Please select the diagram of overlapping circles that best represents your relationship with the activity literati gathering (\textit{Wenren Yaji}). \\
    Options: 1–7 pictorial circles (1 = completely separate … 7 = almost fully overlapping).
    \item \textbf{IOS (post).} Please select the diagram of overlapping circles that best represents your relationship with the activity literati gathering (\textit{Wenren Yaji}). \\
    Options: 1–7 pictorial circles (1 = completely separate … 7 = almost fully overlapping).
\end{itemize}

\subsubsection{Self-Assessment Manikin (SAM): Valence / Arousal / Dominance (pre, img, vid, post)}\label{A.1.3}

\begin{itemize}
    \item Valence — 1–9 (1 = happy, 9 = unhappy)
    \item Arousal — 1–9 (1 = excited, 9 = calm)
    \item Dominance — 1–9 (1 = controlled, 9 = in control)
\end{itemize}

\begin{itemize}
    \item \textbf{SAM (pre, baseline).} Your overall affect right now before any materials.
    \item \textbf{SAM (img, after images).} Your affect right now after viewing the images.
    \item \textbf{SAM (vid, after videos).} Your affect right now after viewing the videos.
    \item \textbf{SAM (post, after all experiences).} Your overall affect right now after completing all experiences.
\end{itemize}

\begin{figure}[h]
    \centering
    \includegraphics[width=0.7\linewidth]{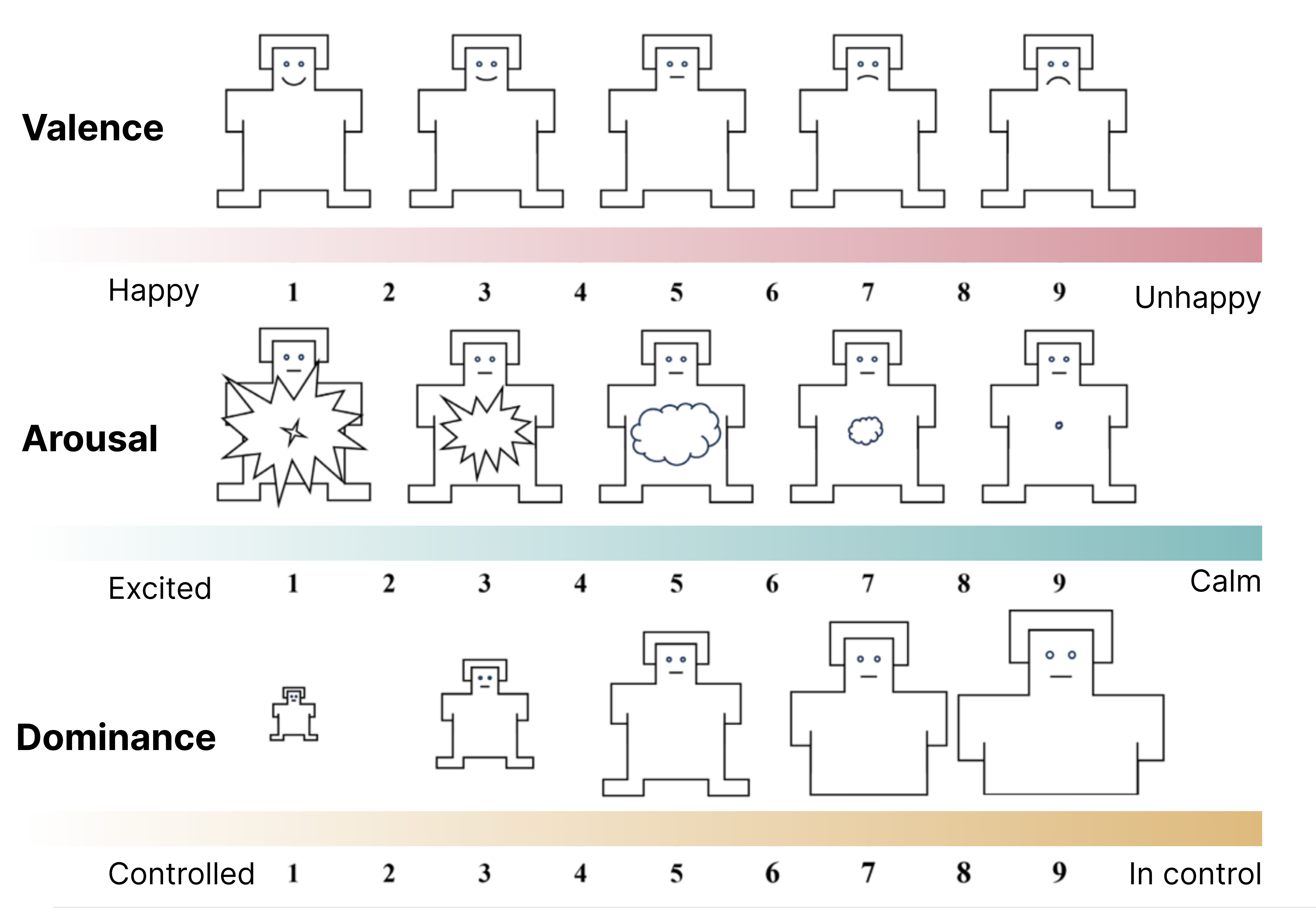}
    \caption{The 9-step SAM scale: from top to bottom: Valence, Arousal, and Dominance.}
    \Description{An illustration of the Self-Assessment Manikin (SAM) scale depicting three affective dimensions: valence, arousal, and dominance. Each dimension is represented by a series of pictorial figures arranged on a 1–9 scale, indicating emotional pleasantness to unpleasantness (valence), excitement to calmness (arousal), and feeling controlled to feeling in control (dominance). The same SAM scales are used at four measurement points: before any materials (pre), after viewing images (img), after viewing videos (vid), and after completing all experiences (post).}
    \label{fig:Appendix2}
\end{figure}

\textit{Options for each dimension: Valence1–9; Arousal1–9; Dominance1–9 (Figure~\ref{fig:Appendix2})}.

\subsubsection{Film-IEQ (Proxy)}\label{A.1.4}

\textit{Response options for all items: 7-point Likert (1 = Strongly disagree … 7 = Strongly agree).}

\textbf{After images:}
\begin{enumerate}
    \item I can understand the main activity or intention conveyed by the image.
    \item The key props/symbols are clearly discernible in the image.
    \item I feel confused or prone to misinterpretations about this image.
\end{enumerate}

\textbf{After videos:}
\begin{enumerate}
    \setcounter{enumi}{3}
    \item I can understand the main activity or intention conveyed by the video.
    \item The key props/symbols are clearly discernible in the video.
    \item I feel confused or prone to misinterpretations about this video.
\end{enumerate}

\textbf{Style ranking (ordering item):}
\begin{enumerate}
    \setcounter{enumi}{6}
    \item Please rank the four visual styles by how much they help you understand literati gathering (\textit{Wenren Yaji}) (1 = most helpful … 4 = least helpful). \\
    \textit{Ink-wash (Xieyi); Fine-brush (Gongbi); Cartoon; Oil.}
\end{enumerate}

\subsubsection{Cultural Resonance (Custom Scale)}\label{A.1.5}

\textit{Response options for all items: 7-point Likert (1 = Strongly disagree … 7 = Strongly agree).}

\textbf{After images:}
\begin{enumerate}
    \item The AI-generated image makes me feel an emotional connection to literati gathering (\textit{Wenren Yaji}).
    \item The AI-generated image helps me sense the cultural meaning intended by the scene.
    \item From the image, I can feel aesthetic interest and appreciation.
\end{enumerate}

\textbf{After videos:}
\begin{enumerate}
    \setcounter{enumi}{3}
    \item The AI-generated video makes me feel an emotional connection to literati gathering (\textit{Wenren Yaji}).
    \item The AI-generated video helps me sense the cultural meaning intended by the scene.
    \item From the video, I can feel aesthetic interest and appreciation.
\end{enumerate}

\textbf{Post (overall):}
\begin{enumerate}
    \setcounter{enumi}{6}
    \item Through the overall experience, I gained a deeper understanding of literati gathering (\textit{Wenren Yaji}).
    \item This form of representation helps me feel closer to traditional culture.
    \item I can sense aesthetic value and appreciation in the scene.
\end{enumerate}

\subsubsection{ITC-SOPI (Proxy)}\label{A.1.6}

\textit{Response options for all items: 7-point Likert (1 = Strongly disagree … 7 = Strongly agree).}

\begin{enumerate}
    \item I felt immersed in the scene, as if I were there. (Spatial Presence)
    \item The details (attire, objects, environment) created a credible atmosphere. (Naturalness)
    \item I could understand the interactions among participants (calligraphy, pitch-pot, Go, poetry). (Engagement / social understanding)
    \item I sensed a social atmosphere rather than an isolated display. (Engagement)
    \item The scene motivated me to continue participating or discussing with others. (Engagement)
    \item From this experience, I could infer the social structure of historical literati gatherings. (Understanding / social framework)
    \item The AI-generated visuals felt natural and credible. (Naturalness)
    \item The media forms (images/videos/interaction) helped me take on a role. (Engagement / role adoption)
    \item I could sense a co-creation / peer-review vibe. (Engagement / co-creation)
\end{enumerate}

\subsection{Semi-Structured Interview Questions}\label{A.2}

\begin{enumerate}
    \item Across the whole experience, which visual or scene impressed you the most? Why did it make you feel this was a literati gathering (\textit{Wenren Yaji}) rather than an ordinary modern get-together?
    \item Were there any parts you found confusing or unclear? How did you infer or guess their meaning at the time?
    \item Among pitch-pot (touhu), calligraphy, Go (Weiqi), and poetry recitation, which activity was presented most clearly in the videos? Which was the least clear, and why?
    \item Among the styles—Fine-brush (Gongbi), Ink-wash (Xieyi), Cartoon and Oil, which do you think best conveys a literati gathering (\textit{Wenren Yaji})?
    \item Was there any style that looked aesthetically pleasing but hindered your understanding of the content? Conversely, was there a style that looked simple yet made comprehension easier?
    \item How did you recognize turn-taking or co-creation in the activities? (For example, from participants’ gestures, gaze, or sound cues.)
    \item Was there a moment that conveyed a strong sense of ritual or solemnity? What produced that feeling?
    \item When you performed calligraphy, pitch-pot, or Go yourself, how did the experience differ from only watching the videos? Did the bodily actions give you any new understanding?
    \item To what extent did the on-site setup and props (e.g., brush, pot, game pieces/board) help you feel embedded in the scene? How would the experience change if these were absent?
    \item During the study, was there a moment when you almost forgot you were in an experiment, as if you were truly present at a historical literati gathering? What was happening then?
    \item Beyond seeing and hearing, would additional sensory cues (e.g., the scent of ink or tea, a light breeze) help you immerse more easily?
    \item When you performed the activities yourself, which information from the videos actually helped (e.g., understanding rules or key techniques more quickly)?
    \item Did you notice differences between what you saw in the videos and what you experienced hands-on? Did those differences make you more confused, or instead feel intrigued?
    \item Did you ever misinterpret something at first but later find that the misunderstanding was interesting or inspiring? Please describe.
    \item If we were to create an immersive literati-gathering experience next, what form would you prefer? If you were the designer, how would you design it?
    \item If the system were to gently prompt you when you might be misinterpreting something, what modality would you prefer (e.g., a brief caption, a short audio cue, a side-by-side comparison clip)? Would you want to control these prompts (on/off)?
    \item (After showing a feature list.) Which three features would you most want to include, and which one would you least want? Why?
    \item If you were explaining ``literati gathering (\textit{Wenren Yaji}'' to a friend by analogy to a present-day activity, what analogy would you use?
    \item Looking back over the experience, what did you like most?
    \item Anything else you would like to add, or any ideas you think we should know?
\end{enumerate}

\subsection{KANO User Requirements Scale}\label{A.3}

\aptLtoX{\begin{enumerate}
    \item[(1)] If the immersive literati-gathering experience included more cultural scenes (e.g., qin/chess-Go/calligraphy/painting; poetry/wine/tea/flowers), how would you feel? 
\end{enumerate}
    \begin{tabular}{ll}
        (a) If this feature is present, your evaluation is: & I like it | It must be that way | I am neutral | I can live with it | I dislike it \\
        (b) If this feature is absent, your evaluation is: & I like it | It must be that way | I am neutral | I can live with it | I dislike it \\
    \end{tabular}
\begin{enumerate}

    \item[(2)] If the experience included virtual character interactions (e.g., poetic peers, literati dialogues), how would you feel? \\
\end{enumerate}

    \begin{tabular}{ll}
        (a) If this feature is present, your evaluation is: & I like it | It must be that way | I am neutral | I can live with it | I dislike it \\
        (b) If this feature is absent, your evaluation is: & I like it | It must be that way | I am neutral | I can live with it | I dislike it \\
    \end{tabular}
\begin{enumerate}

    \item[(3)] If the immersive literati-gathering included multisensory simulations of natural elements (e.g., wind sounds, floral scents, tea aroma), how would you feel? \\
\end{enumerate}
 
   \begin{tabular}{ll}
        (a) If this feature is present, your evaluation is: & I like it | It must be that way | I am neutral | I can live with it | I dislike it \\
        (b) If this feature is absent, your evaluation is: & I like it | It must be that way | I am neutral | I can live with it | I dislike it \\
    \end{tabular}
\begin{enumerate}

    \item[(4)] If the experience added traditional ritual segments (e.g., pitch-pot, tea preparation/whisking, antiphonal poetry), how would you feel? \\
\end{enumerate}

    \begin{tabular}{ll}
        (a) If this feature is present, your evaluation is: & I like it | It must be that way | I am neutral | I can live with it | I dislike it \\
        (b) If this feature is absent, your evaluation is: & I like it | It must be that way | I am neutral | I can live with it | I dislike it \\
    \end{tabular}
\begin{enumerate}

    \item[(5)] If the immersive literati-gathering offered greater spatial freedom (e.g., moving around / choosing different vantage points), how would you feel? \\
\end{enumerate}

    \begin{tabular}{ll}
        (a) If this feature is present, your evaluation is: & I like it | It must be that way | I am neutral | I can live with it | I dislike it \\
        (b) If this feature is absent, your evaluation is: & I like it | It must be that way | I am neutral | I can live with it | I dislike it \\
    \end{tabular}
\begin{enumerate}

    \item[(6)] If the experience added personalization (e.g., choosing attire, seating, or the Four Treasures of the Study—brush, ink, paper, inkstone), how would you feel? \\
\end{enumerate}

    \begin{tabular}{ll}
        (a) If this feature is present, your evaluation is: & I like it | It must be that way | I am neutral | I can live with it | I dislike it \\
        (b) If this feature is absent, your evaluation is: & I like it | It must be that way | I am neutral | I can live with it | I dislike it \\
    \end{tabular}
\begin{enumerate}

    \item[(7)] If the immersive literati-gathering included immersive performances (e.g., integrated poetry–music–dance scenes), how would you feel? \\
    \end{enumerate}

    \begin{tabular}{ll}
        (a) If this feature is present, your evaluation is: & I like it | It must be that way | I am neutral | I can live with it | I dislike it \\
        (b) If this feature is absent, your evaluation is: & I like it | It must be that way | I am neutral | I can live with it | I dislike it \\
    \end{tabular}}{\begin{enumerate}
    \item If the immersive literati-gathering experience included more cultural scenes (e.g., qin/chess-Go/calligraphy/painting; poetry/wine/tea/flowers), how would you feel? 

    \begin{tabular}{ll}
        (a) If this feature is present, your evaluation is: & I like it | It must be that way | I am neutral | I can live with it | I dislike it \\
        (b) If this feature is absent, your evaluation is: & I like it | It must be that way | I am neutral | I can live with it | I dislike it \\
    \end{tabular}

    \item If the experience included virtual character interactions (e.g., poetic peers, literati dialogues), how would you feel? \\

    \begin{tabular}{ll}
        (a) If this feature is present, your evaluation is: & I like it | It must be that way | I am neutral | I can live with it | I dislike it \\
        (b) If this feature is absent, your evaluation is: & I like it | It must be that way | I am neutral | I can live with it | I dislike it \\
    \end{tabular}

    \item If the immersive literati-gathering included multisensory simulations of natural elements (e.g., wind sounds, floral scents, tea aroma), how would you feel? \\

   \begin{tabular}{ll}
        (a) If this feature is present, your evaluation is: & I like it | It must be that way | I am neutral | I can live with it | I dislike it \\
        (b) If this feature is absent, your evaluation is: & I like it | It must be that way | I am neutral | I can live with it | I dislike it \\
    \end{tabular}

    \item If the experience added traditional ritual segments (e.g., pitch-pot, tea preparation/whisking, antiphonal poetry), how would you feel? \\

    \begin{tabular}{ll}
        (a) If this feature is present, your evaluation is: & I like it | It must be that way | I am neutral | I can live with it | I dislike it \\
        (b) If this feature is absent, your evaluation is: & I like it | It must be that way | I am neutral | I can live with it | I dislike it \\
    \end{tabular}

    \item If the immersive literati-gathering offered greater spatial freedom (e.g., moving around / choosing different vantage points), how would you feel? \\

    \begin{tabular}{ll}
        (a) If this feature is present, your evaluation is: & I like it | It must be that way | I am neutral | I can live with it | I dislike it \\
        (b) If this feature is absent, your evaluation is: & I like it | It must be that way | I am neutral | I can live with it | I dislike it \\
    \end{tabular}

    \item If the experience added personalization (e.g., choosing attire, seating, or the Four Treasures of the Study—brush, ink, paper, inkstone), how would you feel? \\

    \begin{tabular}{ll}
        (a) If this feature is present, your evaluation is: & I like it | It must be that way | I am neutral | I can live with it | I dislike it \\
        (b) If this feature is absent, your evaluation is: & I like it | It must be that way | I am neutral | I can live with it | I dislike it \\
    \end{tabular}

    \item If the immersive literati-gathering included immersive performances (e.g., integrated poetry–music–dance scenes), how would you feel? \\

    \begin{tabular}{ll}
        (a) If this feature is present, your evaluation is: & I like it | It must be that way | I am neutral | I can live with it | I dislike it \\
        (b) If this feature is absent, your evaluation is: & I like it | It must be that way | I am neutral | I can live with it | I dislike it \\
    \end{tabular}}
\end{enumerate}

\subsection{Complete Participant Demographics}\label{A.4}
\label{section:complete participant demographics}
\small
\begin{longtable}{cccccc}
\caption{\textbf{Participant demographics.} We retained the raw user-provided data to preserve data integrity. It should be noted that interpretations of some options varied among participants; for example, all users who selected ``High School'' for highest education were current undergraduates. Consequently, no between-group analyses were performed on education levels above high school.} 
\label{tab:Part1}\\
\toprule
No. & Age & Gender & Education & Language & Field \\
\midrule
\endfirsthead
\multicolumn{6}{c}{{\bfseries \tablename\ \thetable{} -- continued from previous page}} \\
\toprule
No. & Age & Gender & Education & Language & Field \\
\midrule
\endhead
\bottomrule
\multicolumn{6}{r}{{Continued on next page}} \\
\endfoot
\bottomrule
\endlastfoot
\multicolumn{6}{c}{\textit{Group 1 (Forward)}} \\
\hline
1 & 18-30 & M & Bachelor & Chinese & Artificial Intelligence \\
2 & 18-30 & M & Bachelor & Chinese & Artificial Intelligence \\
3 & 18-30 & M & High School & Chinese & None \\
4 & 18-30 & M & Master & Chinese & AI-Generated Content \\
5 & 18-30 & M & Master & Chinese & Intangible Cultural Heritage \\
6 & 18-30 & M & Master & Chinese & Visualization Human-Computer Interaction \\
7 & 31-40 & M & Ph.D. & Chinese & Human-Computer Interaction \\
8 & 18-30 & M & Ph.D. & Chinese & Computational Media and Arts \\
9 & 18-30 & F & Bachelor & Chinese & Micro+Artificial Intelligence \\
10 & 18-30 & M & Bachelor & Chinese & Human-Computer Interaction \\
11 & 18-30 & M & Master & Chinese & Economics \\
12 & 18-30 & F & Bachelor & Chinese & Artificial Intelligence \\
13 & 18-30 & F & Bachelor & Chinese & Data Analysis \\
14 & 18-30 & F & Bachelor & Chinese & Smart Manufacturing \\
15 & 18-30 & F & Master & Chinese & Human-Computer Interaction \\
16 & <18 & F & High School & Chinese & None \\
17 & 18-30 & M & Master & Chinese & Psychology \\
18 & 18-30 & F & Master & Chinese & Computational Media and Arts \\
19 & 18-30 & F & Ph.D. & Chinese & Human-Computer Interaction \\
20 & 18-30 & F & Master & Chinese & Computational Media and Arts \\
21 & 18-30 & M & Master & Chinese & Biomaterials \\
22 & <18 & F & Bachelor & Chinese & Artificial Intelligence \\
23 & 18-30 & M & Ph.D. & Chinese & Computer Vision \\
\hline
\multicolumn{6}{c}{\textcolor{black}{\textit{Group 2 (Reverse)}}} \\
\hline
\textcolor{black}{24} & \textcolor{black}{18-30} & \textcolor{black}{M} & \textcolor{black}{Bachelor} & \textcolor{black}{Chinese} & \textcolor{black}{Artificial Intelligence} \\
\textcolor{black}{25} & \textcolor{black}{18-30} & \textcolor{black}{M} & \textcolor{black}{Master} & \textcolor{black}{Chinese} & \textcolor{black}{Calligraphy} \\
\textcolor{black}{26} & \textcolor{black}{18-30} & \textcolor{black}{M} & \textcolor{black}{Bachelor} & \textcolor{black}{Chinese} & \textcolor{black}{Artificial Intelligence} \\
\textcolor{black}{27} & \textcolor{black}{18-30} & \textcolor{black}{M} & \textcolor{black}{Master} & \textcolor{black}{Chinese} & \textcolor{black}{Computer Graphics} \\
\textcolor{black}{28} & \textcolor{black}{18-30} & \textcolor{black}{F} & \textcolor{black}{Master} & \textcolor{black}{Chinese} & \textcolor{black}{Architectural Restoration} \\
\textcolor{black}{29} & \textcolor{black}{18-30} & \textcolor{black}{F} & \textcolor{black}{Master} & \textcolor{black}{Chinese} & \textcolor{black}{Computational Media and Arts} \\
\textcolor{black}{30} & \textcolor{black}{18-30} & \textcolor{black}{M} & \textcolor{black}{Ph.D.} & \textcolor{black}{Chinese} & \textcolor{black}{Robotics and Autonomous Systems Control} \\
\textcolor{black}{31} & \textcolor{black}{18-30} & \textcolor{black}{M} & \textcolor{black}{Ph.D.} & \textcolor{black}{Chinese} & \textcolor{black}{Autonomous Driving} \\
\textcolor{black}{32} & \textcolor{black}{18-30} & \textcolor{black}{M} & \textcolor{black}{Ph.D.} & \textcolor{black}{Chinese} & \textcolor{black}{3D Gaussian Splatting} \\
\textcolor{black}{33} & \textcolor{black}{18-30} & \textcolor{black}{F} & \textcolor{black}{Master} & \textcolor{black}{Chinese} & \textcolor{black}{Intelligent In-Vehicle Human-Computer Interaction} \\
\textcolor{black}{34} & \textcolor{black}{18-30} & \textcolor{black}{F} & \textcolor{black}{Master} & \textcolor{black}{Chinese} & \textcolor{black}{Intangible Cultural Heritage} \\
\textcolor{black}{35} & \textcolor{black}{18-30} & \textcolor{black}{F} & \textcolor{black}{Master} & \textcolor{black}{Chinese} & \textcolor{black}{Serious Games} \\
\textcolor{black}{36} & \textcolor{black}{18-30} & \textcolor{black}{M} & \textcolor{black}{Bachelor} & \textcolor{black}{Chinese} & \textcolor{black}{Triboelectric Nanogenerators} \\
\textcolor{black}{37} & \textcolor{black}{18-30} & \textcolor{black}{M} & \textcolor{black}{Bachelor} & \textcolor{black}{Chinese} & \textcolor{black}{Financial Technology} \\
\textcolor{black}{38} & \textcolor{black}{18-30} & \textcolor{black}{M} & \textcolor{black}{Bachelor} & \textcolor{black}{Chinese} & \textcolor{black}{Sustainable Energy} \\
\hline
\multicolumn{6}{c}{\textcolor{black}{\textit{Group 3 (Non-Native Speaker)}}} \\
\hline
\textcolor{black}{39} & \textcolor{black}{18-30} & \textcolor{black}{M} & \textcolor{black}{Master} & \textcolor{black}{English} & \textcolor{black}{Computer Graphics} \\
\textcolor{black}{40} & \textcolor{black}{18-30} & \textcolor{black}{F} & \textcolor{black}{Bachelor} & \textcolor{black}{English} & \textcolor{black}{Financial Technology} \\
\textcolor{black}{41} & \textcolor{black}{18-30} & \textcolor{black}{M} & \textcolor{black}{High School} & \textcolor{black}{English} & \textcolor{black}{Robotics} \\
\textcolor{black}{42} & \textcolor{black}{18-30} & \textcolor{black}{M} & \textcolor{black}{Master} & \textcolor{black}{English} & \textcolor{black}{3D Generation} \\
\hline
\multicolumn{6}{c}{\textcolor{black}{\textit{Group 4 (Children)}}} \\
\hline
\textcolor{black}{43} & \textcolor{black}{9} & \textcolor{black}{F} & \textcolor{black}{Primary School} & \textcolor{black}{Chinese} & \textcolor{black}{None} \\
\textcolor{black}{44} & \textcolor{black}{7} & \textcolor{black}{M} & \textcolor{black}{Primary School} & \textcolor{black}{Chinese} & \textcolor{black}{None} \\
\textcolor{black}{45} & \textcolor{black}{7} & \textcolor{black}{F} & \textcolor{black}{Primary School} & \textcolor{black}{Chinese} & \textcolor{black}{None} \\
\textcolor{black}{46} & \textcolor{black}{10} & \textcolor{black}{F} & \textcolor{black}{Primary School} & \textcolor{black}{Chinese} & \textcolor{black}{None} \\
\textcolor{black}{47} & \textcolor{black}{9} & \textcolor{black}{F} & \textcolor{black}{Primary School} & \textcolor{black}{Chinese} & \textcolor{black}{None} \\
\textcolor{black}{48} & \textcolor{black}{9} & \textcolor{black}{F} & \textcolor{black}{Primary School} & \textcolor{black}{Chinese} & \textcolor{black}{None} \\
\end{longtable}

\begin{longtable}{cccccc}
\caption{\textbf{Participant experience.} The table records the relevant experience assessments of 48 participants across four dimensions: frequency of cultural contact (Culture Contact), familiarity with literati (Literati Familiarity), frequency of AI media contact (AI Media Contact), and exhibition experience (Exhibition Experience). Each dimension is described using standardized descriptive terms such as "Frequently," "Occasionally," and "Rarely" to quantify the participants' behavioral or cognitive levels.} \label{tab:Part 2}\\
\toprule
No. & Culture Contact & Literati Familiarity & AI Media Contact & Exhibition Experience \\
\midrule
\endfirsthead
\multicolumn{6}{c}{{\bfseries \tablename\ \thetable{} -- continued from previous page}} \\
\toprule
No. & Culture Contact & Literati Familiarity & AI Media Contact & Exhibition Experience \\
\midrule
\endhead
\bottomrule
\multicolumn{6}{r}{{Continued on next page}} \\
\endfoot
\bottomrule
\endlastfoot
1 & Frequently & Somewhat & Daily & Quite experienced \\
2 & Occasionally & Almost un. & Frequently & No experience \\
3 & Rarely & Almost un. & Occasionally & A little bit \\
4 & Frequently & Relatively & Daily & Some experience \\
5 & Occasionally & Somewhat & Very freq. & Quite experienced \\
6 & Frequently & Relatively & Daily & Quite experienced \\
7 & Frequently & Relatively & Daily & Quite experienced \\
8 & Frequently & Somewhat & Frequently & Some experience \\
9 & Occasionally & Somewhat & Frequently & No experience \\
10 & Rarely & Know & Frequently & Some experience \\
11 & Occasionally & Almost un. & Frequently & Some experience \\
12 & Occasionally & Almost un. & Occasionally & A little bit \\
13 & Less freq. & Know & Occasionally & No experience \\
14 & Less freq. & Somewhat & Frequently & No experience \\
15 & Occasionally & Complete un. & Frequently & Very experienced \\
16 & Less freq. & Somewhat & Less freq. & Some experience \\
17 & Occasionally & Almost un. & Occasionally & A little bit \\
18 & Occasionally & Know & Frequently & Some experience \\
19 & Frequently & Know & Frequently & Quite experienced \\
20 & Rarely & Complete un. & Occasionally & Some experience \\
21 & Occasionally & Almost un. & Frequently & A little bit \\
22 & Frequently & Relatively & Occasionally & No experience \\
23 & Frequently & Know & Less freq. & Quite experienced \\
24 & Occasionally & Know & Frequently & Some experience \\
\textcolor{black}{25} & \textcolor{black}{Frequently} & \textcolor{black}{Almost un.} & \textcolor{black}{Frequently} & \textcolor{black}{A little bit} \\
\textcolor{black}{26} & \textcolor{black}{Frequently} & \textcolor{black}{Know} & \textcolor{black}{Frequently} & \textcolor{black}{Some experience} \\
\textcolor{black}{27} & \textcolor{black}{Frequently} & \textcolor{black}{Know} & \textcolor{black}{Frequently} & \textcolor{black}{A little bit} \\
\textcolor{black}{28} & \textcolor{black}{Less freq.} & \textcolor{black}{Almost un.} & \textcolor{black}{Less freq.} & \textcolor{black}{Quite experienced} \\
\textcolor{black}{29} & \textcolor{black}{Less freq.} & \textcolor{black}{Somewhat} & \textcolor{black}{Frequently} & \textcolor{black}{Some experience} \\
\textcolor{black}{30} & \textcolor{black}{Less freq.} & \textcolor{black}{Almost un.} & \textcolor{black}{Frequently} & \textcolor{black}{Some experience} \\
\textcolor{black}{31} & \textcolor{black}{Occasionally} & \textcolor{black}{Relatively} & \textcolor{black}{Frequently} & \textcolor{black}{A little bit} \\
\textcolor{black}{32} & \textcolor{black}{Occasionally} & \textcolor{black}{Relatively} & \textcolor{black}{Less freq.} & \textcolor{black}{Some experience} \\
\textcolor{black}{33} & \textcolor{black}{Frequently} & \textcolor{black}{Relatively} & \textcolor{black}{Frequently} & \textcolor{black}{Quite experienced} \\
\textcolor{black}{34} & \textcolor{black}{Less freq.} & \textcolor{black}{Know} & \textcolor{black}{Frequently} & \textcolor{black}{Some experience} \\
\textcolor{black}{35} & \textcolor{black}{Less freq.} & \textcolor{black}{Know} & \textcolor{black}{Frequently} & \textcolor{black}{Some experience} \\
\textcolor{black}{36} & \textcolor{black}{Occasionally} & \textcolor{black}{Relatively} & \textcolor{black}{Frequently} & \textcolor{black}{No experience} \\
\textcolor{black}{37} & \textcolor{black}{Rarely} & \textcolor{black}{Almost un.} & \textcolor{black}{Less freq.} & \textcolor{black}{Some experience} \\
\textcolor{black}{38} & \textcolor{black}{Less freq.} & \textcolor{black}{Somewhat} & \textcolor{black}{Rarely} & \textcolor{black}{No experience} \\
\textcolor{black}{39} & \textcolor{black}{Occasionally} & \textcolor{black}{Somewhat} & \textcolor{black}{Frequently} & \textcolor{black}{No experience} \\
\textcolor{black}{40} & \textcolor{black}{Rarely} & \textcolor{black}{Somewhat} & \textcolor{black}{Frequently} & \textcolor{black}{Some experience} \\
\textcolor{black}{41} & \textcolor{black}{Occasionally} & \textcolor{black}{Almost un.} & \textcolor{black}{Occasionally} & \textcolor{black}{A little bit} \\
\textcolor{black}{42} & \textcolor{black}{Rarely} & \textcolor{black}{Somewhat} & \textcolor{black}{Occasionally} & \textcolor{black}{Completely un.} \\
\textcolor{black}{43} & \textcolor{black}{Rarely} & \textcolor{black}{Almost un.} & \textcolor{black}{Daily} & \textcolor{black}{Some experience} \\
\textcolor{black}{44} & \textcolor{black}{Occasionally} & \textcolor{black}{Complete un.} & \textcolor{black}{Less freq.} & \textcolor{black}{A little bit} \\
\textcolor{black}{45} & \textcolor{black}{Occasionally} & \textcolor{black}{Complete un.} & \textcolor{black}{Frequently} & \textcolor{black}{No experience} \\
\textcolor{black}{46} & \textcolor{black}{Less freq.} & \textcolor{black}{Relatively} & \textcolor{black}{Frequently} & \textcolor{black}{Some experience} \\
\textcolor{black}{47} & \textcolor{black}{Occasionally} & \textcolor{black}{Complete un.} & \textcolor{black}{Rarely} & \textcolor{black}{No experience} \\
\textcolor{black}{48} & \textcolor{black}{Occasionally} & \textcolor{black}{Complete un.} & \textcolor{black}{Rarely} & \textcolor{black}{No experience} \\
\end{longtable}

\end{document}